\documentclass[showpacs,12pt,preprint,preprintnumbers,nofootinbib,groupedaddress,amsmath,amssymb]{revtex4}

\usepackage{graphicx}% Include figure files
\usepackage{dcolumn}% Align table columns on decimal point
\usepackage{bm}% bold math
\usepackage{amssymb}
\usepackage{amsmath}
\usepackage{epsfig} 

\def\beq{\begin{equation}}
\def\eeq{\end{equation}}
\def\bea{\begin{array}}
\def\eea{\end{array}}
\def\be{\begin{equation}}
\def\ee{\end{equation}}
\def\ba{\begin{eqnarray}}
\def\ea{\end{eqnarray}}

\def\to{\rightarrow}

\def\[{\left[}
\def\]{\right]}
\def\({\left(}
\def\){\right)}

\def\sm0{{\widetilde{m}_0}}

\def\U1em{{U(1)_{\rm em}}}
\def\to{\rightarrow}

\def\sq2{\sqrt{2}}

\def\ee{e^+e^-}

\def\End{\end{document}}
\newcommand{\gsim}{\mbox{ \raisebox{-1.0ex}{$\stackrel{\textstyle >}
{\textstyle \sim}$ }}}
\newcommand{\lsim}{\mbox{ \raisebox{-1.0ex}{$\stackrel{\textstyle <}
{\textstyle \sim}$ }}}

\def\fsl#1{\setbox0=\hbox{$#1$}                 % set a box for #1 
   \dimen0=\wd0                                 % and get its size
   \setbox1=\hbox{/} \dimen1=\wd1               % get size of /
   \ifdim\dimen0>\dimen1                        % #1 is bigger
      \rlap{\hbox to \dimen0{\hfil/\hfil}}      % so center / in box
      #1                                        % and print #1
   \else                                        % / is bigger
      \rlap{\hbox to \dimen1{\hfil$#1$\hfil}}   % so center #1
      /                                         % and print /
   \fi}

\begin{document}

\title{A Model of TeV Scale Physics for Neutrino Mass, Dark Matter and
       Baryon Asymmetry and its Phenomenology}% and its Phenomenology}
\author{%
{\sc Mayumi Aoki\,$^1$, Shinya Kanemura\,$^2$,
   and Osamu Seto\,$^3$}
}
\affiliation{%
%\address{\vspace*{5mm}
\vspace*{2mm} 
$^1$Department of Physics,~Tohoku University,~Aramaki, Aoba, Sendai,~Miyagi 980-8578, Japan\\
$^2$Department of Physics, University of Toyama, 3190 Gofuku, Toyama 930-8555, Japan\\ 
$^3$William I. Fine Theoretical Physics Institute, University of Minnesota,
     Minneapolis, MN 55455, USA\footnote{Address after April 2009: 
 Department of Architecture and Building Engineering, Hokkai-Gakuen University, Sapporo 062-8605, Japan}}
%\maketitle

\preprint{TU-846, UT-HET 027, FTPI-MINN-09/16, UMN-TH-2744/09}

%\vspace*{5mm} 
\begin{abstract}
 We discuss some details of the model proposed in Ref.~\cite{aks-prl},
 in which  neutrino oscillation, dark matter, and baryon asymmetry of 
 the Universe would be simultaneously explained by the TeV-scale physics
 without introducing very high mass scales.
 An exact discrete $Z_2$ symmetry is introduced, under which
 new particle contents (a real singlet scalar field, a pair of 
 charged singlet scalar fields and TeV-scale right-handed neutrinos)
 are assigned to have odd quantum number, whereas ordinary gauge
 fields, quarks and leptons, and two Higgs doublets are even.  
 Tiny neutrino masses are generated at the three loop level due to
 the exact $Z_2$ symmetry, by which stability of the dark
 matter candidate is also guaranteed. The extra Higgs doublet is
 required not only for the tiny neutrino masses but also for
 successful electroweak baryogenesis.
 We  discuss phenomenological properties of the model, 
and find that there are successful scenarios in which
above three problems are solved simultaneously 
under the constraint from current experimental data.
 We then discuss predictions in such scenarios
 at ongoing and future experiments.
 It turns out that the model provides discriminative predictions
 especially in Higgs physics and dark matter physics, 
 so that it is testable in near future.
\pacs{\, 14.60.Pq, 14.60.St, 14.80.Cp, 12.60.Fr 
\hfill   ~~ [\today] }
\end{abstract}

\maketitle

\setcounter{footnote}{0}
\renewcommand{\thefootnote}{\arabic{footnote}}

\section{Introduction} 

%\subsection{The reason for beyond the SM}

While the standard model (SM) for elementary particles has been
successful in describing high energy phenomena at colliders, 
today we have definite motivation to consider a model beyond the SM.
First of all, observed data for neutrino oscillation indicate that neutrinos
have tiny masses and mix with each other~\cite{lep-data}. 
Second, cosmological data have revealed that the density of 
dark energy and dark matter (DM) in the Universe dominates 
that of baryonic matter~\cite{wimp}. The essence of DM would be 
weakly interacting massive particles (WIMPs). 
Finally, asymmetry of matter and anti-matter in our Universe has been addressed
for a long time as a serious problem regarding existence of ourselves~\cite{sakharov}.
They are all beyond the scope of the SM, 
so that a new model is required to explain these phenomena.

%
%\subsection{Scenarios with very large mass hierarchy}
%
%\indent
A simple scenario to generate tiny  neutrino masses ($m_\nu$) would be 
based on the see-saw mechanism with heavy right-handed (RH) Majorana neutrinos~\cite{see-saw}.
Then, $m_\nu$ can be described as $m_\nu \simeq m_D^2/M_R$, where $M_R$ ($\sim
10^{13-16}$ GeV) is the Majorana mass of RH neutrinos and 
$m_D^{}$ is the Dirac mass of at most the electroweak scale.  
This scenario would be compatible with the framework with
large mass scales like grand unification.
The heavy RH Majorana neutrinos would generate lepton
number asymmetry in their CP violating decays that could be transfered into baryon
asymmetry~\cite{FY}. 
In a supersymmetric model with such heavy RH neutrinos, 
one might find a DM candidate of lightest supersymmetric 
particle~\cite{Jungman:1995df}.
Introduction of such large scales,
however, causes a problem of hierarchy. 
In addition, the decoupling theorem~\cite{dec-theorem}
makes it far from experimental tests.

An alternative approach
is a quantum mechanical generation of small neutrino masses.
The original idea of radiatively generating neutrino masses
due to TeV-scale physics has been proposed
by Zee~\cite{zee}.
Adding an isospin doublet scalar field and a charged singlet field to the SM, the neutrino masses
are generated at the one-loop level.
Phenomenology in the Zee model has been studied
in Ref.~\cite{zee-ph2, zee-ph}.
Although the original Zee model has been excluded
by the neutrino data, lots of studies for its extensions and variations
have been proposed~\cite{zee-ex1}.
Another possibility for generating neutrino masses via the new scalar particles is the Zee-Babu model~\cite{ZeeBabu,ZeeBabu-ph},
where the neutrino masses arise at the two-loop level.
Some extensions of these models are discussed
for the purpose of the explanation
for DM and baryon asymmetry~\cite{Zee-LG,ZeeBabu-LG} .
The extension with a TeV-scale RH neutrino
has been discussed in Ref.~\cite{knt}, where the neutrino
masses are generated at the three-loop due to the exact $Z_2$
symmetry which forbids the Dirac neutrino mass at the tree level unlike the
tree-level seesaw mechanism mentioned above,
and the $Z_2$-odd RH neutrino is a candidate of DM.
This model has been extended with two RH neutrinos for the description of the
neutrino data~\cite{kingman-seto},
which however cannot include a mechanism of baryon number generation.
The idea of simultaneous explanation for radiative generation of
neutrino masses and stability of DM by introducing various
exact discrete symmetries with right-handed neutrinos has been used
in several models~\cite{NRandDM}.
Some other models including mechanisms for baryon asymmetry have been considered
in the leptogenesis scenario~\cite{NRandLG} and also in the scenario of
electroweak baryogenesis~\cite{NRandEWBG,aks-prl}.

%
%\subsection{Our model}
%
%\subsubsection{what is new with our model}
%
\indent
In this paper, we investigate some details of a model proposed in
Ref.~\cite{aks-prl}, in which neutrino oscillation, origin of DM 
and baryon asymmetry would be simultaneously explained by the TeV-scale physics. 
As we try to built a renormalizable TeV scale model to explain
these phenomena simultaneously without fine-tuning, 
we do not impose unnatural hierarchy among the mass scales.
The model contains an extended Higgs sector with TeV-scale RH neutrinos
in addition to the SM particle contents.
Tiny masses of left-handed (LH) neutrinos are generated at the three-loop
level under an exact $Z_2$ symmetry. 
The lightest neutral $Z_2$-odd state is a candidate of DM. 
Baryon asymmetry can be generated at the electroweak phase transition
(EWPT) by the non-decoupling property~\cite{ewbg-thdm2} and
additional CP violating phases in the Higgs sector~\cite{ewbg-thdm,ewbg-thdm3}.
In this framework, a successful model can be built 
without contradiction of the current data.
Notice that in this model we do not intend to solve the so-called hierarchy problem.
The model predicts quadratic divergences in quantum corrections to
masses of the scalar bosons as in the SM. This model, therefore,  has to 
be considered as an effective theory below the cut-off scale of a
more fundamental theory around at most $10$~TeV,
below which the self-coupling constants for additional scalar bosons
do not become larger than the acceptable values for a perturbation calculation.
 
We show that there are several possible scenarios, in which 
the data for neutrinos, lepton flavor violation as well as the WMAP
data are satisfied. The collider data from experiments at LEP,
Fermilab Tevatron and B factories at KEK and SLAC are also taken into account.
In addition to the original scenario discussed in Ref~\cite{aks-prl},
we mention another scenario where the recent data from PAMELA and ATIC
would also be included with the relatively heavy DM  candidate.
Furthermore, we also discuss the other scenario where the anomaly in the
data from DAMA$/$LIBRA would be explained by a light DM candidate whose
mass is around 5 GeV.
It turns out that in these scenarios all the masses of additional physical particles
are between ${\cal O}(10)$ GeV and ${\cal O}(1)$ TeV. 
We find that the model has discriminative features in Higgs phenomenology,
lepton flavor physics and DM physics,
so that it is testable at current and future experiments.

%This paper is organized as follows.
In Sec.~II, we define the model with introducing new particle entries and
symmetries, 
and discuss the physical states and their masses and coupling constants.
In Sec.~III, we calculate the neutrino mass matrix in our model,
which is induced at the three loop level,
and discuss parameter sets in which all the neutrino data are reproduced
under current experimental bounds.
Sec.~IV is devoted to the discussion on the possibility that the
lightest $Z_2$ odd scalar boson is a candidate of DM.
The thermal relic abundance is evaluated in several parameter sets, and
implication for the physics at direct and indirect search experiments
are also discussed. In Sec.~V, we study the allowed region where the
strong first order electroweak phase transition is realized. This is
required for a successful scenario of electroweak baryogenesis.
In Sec.~VI, we summarize the constraints from the current
experimental data on the model, and discuss phenomenological
predictions  at ongoing and future collider experiments and at
direct/indirect searches for DM.
Some formulas are summarized in the Appendices.

\section{Model} 

\subsection{Particle contents and symmetries}

In addition to the {\it known} SM fields, new particle entries in our
model are
\begin{eqnarray}
   \Phi_1, \Phi_2, S^\pm, \eta, N_R^\alpha, 
 \end{eqnarray}
where $\Phi_1$, $\Phi_2$ are scalar isospin doublet fields with the 
hypercharge $1/2$, $S^\pm$ are charged isospin singlet fields,
$\eta$ is a real scalar singlet field, and
$N_R^\alpha$ is the $\alpha$-th generation isospin-singlet RH
neutrinos. It turns out that at least two generations are 
necessary for $N_R^\alpha$ to reproduce the neutrino data.
In the following, we mainly consider the minimum model with two
generation $N_R^\alpha$ ($\alpha=1,2$).
We give a comment for the case of more than three generations later.
We here only note that the mass scale of new particles derived in the following sections 
is not so sensitive against the number of $N_R$ generation.
This means that the model has high predictability for the mass scale of new particles.

In order to generate tiny neutrino masses in the three loop level, 
in other words to forbid the tree-level Dirac neutrino mass term, 
and at the same time in order to have a stable DM candidate,
we impose a new (exact) $Z_2$ symmetry as in Ref.~\cite{knt}, which we
refer to as $Z_2$. 
We assign the $Z_2$ odd charge to $N_R^\alpha$, $S^\pm$ and $\eta$, while 
ordinary gauge fields, quarks and leptons and Higgs doublets are $Z_2$ even.

It has been well known that introduction of the extra Higgs doublet
causes a problem of dangerous flavor changing neutral current (FCNC).
To avoid FCNC in a natural way, we further introduce the
(softly-broken) $Z_2$ symmetry in the model,
which can be softly broken~\cite{glashow-weinberg} in the Higgs potential.
We refer to this symmetry as the $\tilde{Z}_2$ symmetry in this paper.
Under the $\tilde{Z}_2$ symmetry, there can be four independent types of the
Yukawa interaction, depending on the assignment of $\tilde{Z}_2$ charges
for quarks and leptons. 
From a phenomenological reason discussed later, we employ the so-called
Type-X Yukawa coupling where $\tilde{Z}_2$ charges are assigned
such that only $\Phi_1$ couples to leptons whereas $\Phi_2$ does
to quarks~\cite{barger,grossman,typeX,typeX2}.
We summarize the particle properties under 
$Z_2$ and $\tilde{Z}_2$ in TABLE~\ref{discrete}.
\begin{table}
\begin{center}
  \begin{tabular}{c|ccccc|cc|ccc}
   \hline
   & $Q^i$ & $u_R^{i}$ & $d_R^{i}$ & $L^i$ & $e_R^i$ & $\Phi_1$ & $\Phi_2$ & $S^\pm$ &
    $\eta$ & $N_{R}^{\alpha}$ \\\hline
$Z_2\frac{}{}$                ({\rm exact}) & $+$ & $+$ & $+$ & $+$ & $+$ & $+$ & $+$ & $-$ & $-$ & $-$ \\ \hline  
$\tilde{Z}_2\frac{}{}$ ({\rm softly\hspace{1mm}broken})& $+$ & $-$ & $-$ & $+$ &
                       $+$ & $+$ & $-$ & $+$ & $-$ & $+$ \\\hline
   \end{tabular}
\end{center}
  \caption{Particle properties under the discrete symmetries.
 %$Q^i$ and
 %$L^i$ are $i$-th generation quark and lepton LH doublets, respectively
 }
  \label{discrete}
 \end{table}
 
Under these discrete symmetries, the Yukawa interaction is given by 
\begin{eqnarray}
 {\cal L}_Y
%  &=& -  \sum_{i,j=1}^3 y_{\rm \ell}^{ij}  \overline{L}^i
%  \Phi_1 \ell_R^j
%  - y_{\rm u}^{ij}  \overline{Q}^i
%  \tilde{\Phi}_2 u_R^j - y_{\rm d}^{ij}  \overline{Q}^i
%  \Phi_2 d_R^j + {\rm h.c.} \nonumber\\
   = - y_{\ell_i}  \overline{L}^i \Phi_1 \ell_R^i
     - y_{u_i}  \overline{Q}^i \tilde{\Phi}_2 u_R^i
     - y_{d_i}  \overline{Q}^i \Phi_2 d_R^i + {\rm h.c.}, \label{eq:yukawa1}
\end{eqnarray}
where $Q^i$ ($L^i$) is the ordinary $i$-th generation LH quark (lepton)
doublet, and $u_R^i$ and $d_R^i$ ($e_R^i$) are RH-singlet up- and
down-type quarks (charged leptons), respectively.   
Notice that the Type-X Yukawa coupling defined in Eq.~(\ref{eq:yukawa1})~\cite{barger,grossman,typeX,typeX2}
is different from that in the Type-I or Type-II two-Higgs-doublet model (THDM)~\cite{hhg}.
Our $\tilde{Z}_2$ charge assignment for quarks is the same as that in Type I,
but that for leptons is the same as of Type II. The charged Higgs boson couplings to leptons are
multiplied by $\tan\beta$, while those to quarks are by $\cot\beta$
in a universal way, where $\tan\beta=\langle \Phi_2^0
\rangle/\langle \Phi_1^0 \rangle$. %\\
Therefore, phenomenology of the Higgs sector in our model is completely
different from that in Type I and Type II.

The scalar potential is then given by
\begin{eqnarray}
&& V = -\mu_1^2 |\Phi_1|^2 -\mu_2^2 |\Phi_2|^2 - (\mu_{12}^2
  \Phi_1^\dagger \Phi_2 + {\rm h.c.}) \nonumber \\
&&+ \lambda_1|\Phi_1|^4   +
  \lambda_2|\Phi_2|^4 + \lambda_3|\Phi_1|^2|\Phi_2|^2 +\lambda_4 |\Phi_1^\dagger \Phi_2|^2 \  +
   \left\{ \frac{\lambda_5}{2} (\Phi_1^\dagger
    \Phi_2)^2 + {\rm h.c.} \right\} \nonumber\\ 
 && +\mu_s^2 |S|^2 + \lambda_s |S|^4 + \frac{1}{2}\mu_\eta^2 \eta^2 +
  \lambda_\eta \eta^4 + \xi |S|^2 \frac{\eta^2}{2}
 + \sum_{a=1}^2 \left(\rho_a |\Phi_a|^2|S|^2 + \sigma_a |\Phi_a|^2
  \frac{\eta^2}{2}\right) \nonumber\\
 && +\sum_{a,b=1}^2\left\{ \kappa \,\,\epsilon_{ab} (\Phi^c_a)^\dagger
                    \Phi_b S^- \eta + {\rm h.c.}\right\},
 \end{eqnarray}
 where $\epsilon_{ab}$ is the anti-symmetric tensor with $\epsilon_{12}=1$.
The mass term and the interaction for $N_R^\alpha$ are given by 
\begin{eqnarray}
 {\cal L}_Y \!= \! \sum_{\alpha=1}^2\!\left\{ \!\frac{1}{2}m_{N_R^\alpha}^{} \overline{{N_R^\alpha}^c} N_R^\alpha
                 -  h_i^\alpha \overline{(e_R^i)^c}
                   N_R^\alpha S^+\! + {\rm h.c.}\!\right\}.
\end{eqnarray} 

The parameters $\mu_{12}^2$, $\lambda_5$ and $\kappa$ as well as
$h_i^\alpha$ are generally complex.
The phases of $\lambda_5$ and $\kappa$ can be eliminated by re-phasing 
$S^\pm$ and $\Phi_1$. The remaining phase of $\mu_{12}^2$ is physical
and causes CP violation in the Higgs sector,
which is necessary
for generating baryon asymmetry at the EWPT~\cite{ewbg-thdm,ewbg-thdm3}.
Although the CP violating phase is crucial for successful baryogenesis,
it does not much affect in the following discussions on neutrino masses, 
 DM and the strong first order EWPT required for electroweak baryogenesis.
Thus, in the following, we neglect the phase of $\mu_{12}^2$ (and $h_i^\alpha$)
for simplicity.
We later give a comment on how the phenomenology could
change with non-zero CP-violating phase.\\

\subsection{Higgs states and coupling constants}

Because $Z_2$ is exact, $Z_2$ even and odd fields cannot mix.
The Higgs doublet fields $\Phi_i$ ($i=1,2$) can be parameterized as
\begin{align}
\Phi_i=\begin{pmatrix}\omega_i^+\\\frac1{\sqrt2}(v_i+h_i+i\,z_i)
\end{pmatrix},
\end{align}
where $v_i$ are vacuum expectation values and satisfy 
$\sqrt{v_1^2+v_2^2}=v$ ($\simeq 246$ GeV), and 
 $\tan\beta=v_2/v_1$.
The $Z_2$ even states are diagonalized in mass as 
a usual THDM by introducing the mixing angles $\alpha$ and $\beta$, 
where $\alpha$ is that between CP even neutral Higgs bosons~\cite{hhg};
 \begin{eqnarray}
  \left[ \begin{array}{c}
          w^\pm_1 \\ 
          w^\pm_2 \\
         \end{array}\right]
   = 
  \left[ \begin{array}{cc}
          \cos\beta & - \sin\beta\\
          \sin\beta & \cos\beta\\
         \end{array} \right]
  \left[ \begin{array}{c}
          w^\pm \\
          H^\pm \\
         \end{array}
  \right], \hspace{1cm}
% \end{eqnarray}
%
%\begin{eqnarray}
  \left[ \begin{array}{c}
          z_1 \\
          z_2 \\
         \end{array}\right]
   = 
  \left[ \begin{array}{cc}
          \cos\beta & - \sin\beta \\
          \sin\beta & \cos\beta \\
         \end{array}\right]
  \left[ \begin{array}{c}
          z \\
          A \\
         \end{array}\right],
 \end{eqnarray}
and
 \begin{eqnarray}
  \left[ \begin{array}{c}
          h_1  \\
          h_2  \\
         \end{array}\right]
   = 
  \left[ \begin{array}{cc}
          \cos\alpha & - \sin\alpha\\
          \sin\alpha & \cos\alpha\\
         \end{array}\right]
  \left[ \begin{array}{c}
          H \\
          h \\
         \end{array}\right],
 \end{eqnarray}
where $w^\pm$ and $z$ are Nambu-Goldstone bosons eaten by
$W^\pm_L$ and $Z_L$, $H^\pm$ and $A$ are charged and
CP-odd scalar states. CP-odd state can be a mass eigenstate only when
CP is conserved in the Higgs sector.
When CP is conserved $h$ and $H$ are mass eigenstates.
Consequently, the $Z_2$ even physical scalar states are
two CP-even ($h$ and $H$), a CP-odd ($A$) and charged ($H^\pm$) states,
like in usual THDMs.
Throughout this paper, $\sin(\beta-\alpha)=1$ is taken, under which 
we define $h$ and $H$ such that $h$ is the SM-like Higgs boson
\footnote{Notice that this is different from the definition
that $h$ is always the lighter CP even Higgs boson when the SM-like
Higgs is heavier than the other one.}.

The mass formulas for the scalar fields are calculated as\footnote{For
the expressions in the case of $\sin(\beta-\alpha)=1$, see Appendix A.} 
\begin{eqnarray}
  m_h^2 &=&  \sin^2(\alpha-\beta) M_{11}^2+\sin 2(\alpha-\beta) M_{12}^2
   +\cos^2(\alpha-\beta) M_{22}^2, \\
  m_H^2 &=&  \cos^2(\alpha-\beta) M_{11}^2+\sin 2(\alpha-\beta) M_{12}^2
   +\sin^2(\alpha-\beta) M_{22}^2, \\
  m_{H^\pm}^2 &=& M^2-\frac{\lambda_4+\lambda_5}{2} v^2, \\ 
  m_A^2 &=& M^2 -\lambda_5 v^2, \\
  m_{S^\pm}^2 &=& \mu_S^2 + (\rho_1 \cos^2\beta + \rho_2 \sin^2\beta)\frac{v^2}{2}, \\
  m_{\eta}^2 &=& \mu_\eta^2 + (\sigma_1 \cos^2\beta + \sigma_2
   \sin^2\beta)\frac{v^2}{2}, 
\end{eqnarray}
where the mass matrix elements for the CP-even bosons are given by 
\begin{eqnarray}
 M_{11}^2 &=& 
     2(\lambda_1\cos^4\beta+\lambda_2\sin^4\beta+\lambda\cos^2\beta\sin^2\beta)   v^2,\\
M_{12}^2&=&M_{21}^2 = (-2\lambda_1\cos^2\beta+2\lambda_2\sin^2\beta+\lambda\cos 2\beta)
   \sin\beta\cos\beta v^2,           \\
M_{22}^2&=&       M^2+\frac{1}{4} (\lambda_1+\lambda_2-\lambda)
       (1-\cos 4\beta) v^2, 
\end{eqnarray}
where $\lambda=\lambda_3+\lambda_4+\lambda_5$. The parameter $M$
relates to the invaiant mass scale $\mu_{12}$ in the Higgs
potential by $M=|\mu_{12}|/(\sin\beta\cos\beta)$ and has the meanings of
the soft breaking parameter for the $\tilde{Z}_2$ symmetry. 
 
In terms of mass eigenstates, the scalar interaction terms are given by 
\begin{eqnarray}
&& {\cal L}_\rho = \nonumber\\
&&- 
    (\rho_1 \cos^2\beta + \rho_2 \sin^2\beta) \left(\omega^-\omega^+ +
                                               \frac{z^2}{2}
                                              \right)S^+S^-
-    (\rho_1 \sin^2\beta + \rho_2 \cos^2\beta) \left(H^+H^- +
                                                \frac{A^2}{2}
                                               \right)S^+S^- \nonumber\\
&&- \sin\beta\cos\beta (-\rho_1 + \rho_2) (\omega^+ H^- + \omega^- H^+ + z
A)S^+S^- \nonumber\\
&&- (\rho_1 \cos\alpha\cos\beta+\rho_2\sin\alpha\sin\beta)v H S^+S^-
- (- \rho_1 \sin\alpha\cos\beta+\rho_2\cos\alpha\sin\beta)v h S^+S^-\nonumber\\
&&- \frac{1}{2} (\rho_1 \cos^2\alpha + \rho_2 \sin^2\alpha) H H S^+S^-
- \frac{1}{2} (\rho_1 \sin^2\alpha + \rho_2 \cos^2\alpha) h h S^+S^-\nonumber\\
&&-  \cos\alpha\sin\alpha(- \rho_1+ \rho_2) h H S^+S^-, 
 \end{eqnarray}
 \begin{eqnarray}
&& {\cal L}_\sigma = \nonumber\\
&&  -     (\sigma_1 \cos^2\beta + \sigma_2 \sin^2\beta) \left(\omega^-\omega^+ +
                                               \frac{z^2}{2}
                                              \right) \frac{\eta^2}{2}
-    (\sigma_1 \sin^2\beta + \sigma_2 \cos^2\beta) \left(H^+H^- +
                                                \frac{A^2}{2}
                                               \right)\frac{\eta^2}{2} \nonumber\\
&&- \sin\beta\cos\beta (-\sigma_1 + \sigma_2) (\omega^+ H^- + \omega^- H^+ + z
A)\frac{\eta^2}{2} \nonumber\\
&&- (\sigma_1 \cos\alpha\cos\beta+\sigma_2\sin\alpha\sin\beta)v H \frac{\eta^2}{2}
- (- \sigma_1 \sin\alpha\cos\beta+\sigma_2\cos\alpha\sin\beta)v h \frac{\eta^2}{2}\nonumber\\
&&- \frac{1}{2} (\sigma_1 \cos^2\alpha + \sigma_2 \sin^2\alpha) H H \frac{\eta^2}{2}
- \frac{1}{2} (\sigma_1 \sin^2\alpha + \sigma_2 \cos^2\alpha) h h \frac{\eta^2}{2}\nonumber\\
&&-  \cos\alpha\sin\alpha(- \sigma_1+ \sigma_2) h H \frac{\eta^2}{2}, 
 \end{eqnarray}
and 
\begin{eqnarray}
 {\cal L}_\kappa &=& -
   \sqrt{2} \kappa \left[ v H^+
-\sin(\alpha-\beta) H \omega^+
-\cos(\alpha-\beta) h \omega^+ \right. \nonumber\\
&&\left.+\cos(\alpha-\beta) H H^+
-\sin(\alpha-\beta) h H^+ \right] S^- \eta + {\rm h.c.} \,.
 \end{eqnarray}
 The Yukawa interactions with quarks and leptons in
Eq.~(\ref{eq:yukawa1}) can then be written as
\begin{eqnarray}
 {\cal L}_Y^{\rm Quarks} &=& - m_{u^i} \bar u^i u^i - \frac{m_{u^i} \sin\alpha}{v \sin\beta} H \bar u^i
  u^i - \frac{m_{u^i} \cos\alpha}{v \sin\beta} h \bar u^i u^i  
\nonumber\\
&&
 - m_{d^i} \bar d^i d^i - \frac{m_{d^i}
          \sin\alpha}{v\sin\beta} H \bar d^i d^i 
  - \frac{m_{d^i} \cos\alpha}{v\sin\beta} h \bar d^i d^i
\nonumber\\
&&
 + \frac{m_{u^i}}{v} z \bar u^i i \gamma_5 u^i - \frac{m_{u^i}}{v}\cot\beta A
  \bar u^i i \gamma_5 u^i
%  \nonumber\\
% &&
  +\frac{m_{d^i}}{v} z \bar d^i i \gamma_5 d^i - \frac{m_{d^i}}{v} \cot\beta A \bar
  d^i i \gamma_5 d^i  \nonumber\\
 && + \omega^- \bar d^i \left[ \frac{\sqrt{2}m_{u^i}}{v}
                       \left(\frac{1+\gamma_5}{2}\right)-\frac{\sqrt{2}m_{d^i}}{v}
                       \left(\frac{1-\gamma_5}{2}\right)\right] u^i + {\rm h.c.}  \nonumber\\
 && + H^- \bar d^i \left[ \frac{\sqrt{2} m_{u^i}\cot\beta}{v} 
                       \left(\frac{1+\gamma_5}{2}\right)-\frac{\sqrt{2}m_{d^i}\cot\beta}{v}
                       \left(\frac{1-\gamma_5}{2}\right)\right] u^i + {\rm h.c.}  \,.
\end{eqnarray}
 \begin{eqnarray}
{\cal L}_Y^{\rm Leptons}&=&
    +\frac{\sqrt{2}m_{\ell^i}}{v}\bar \nu^i \left(\frac{1+\gamma_5}{2}\right) \ell^i
    \omega^+
%    +
%    {\rm h.c.}
%\nonumber\\
% &&
    -\frac{\sqrt{2}m_{\ell^i} \tan\beta}{v} \bar \nu^i \left(\frac{1+\gamma_5}{2}\right) \ell^i
    H^++{\rm h.c.}
    \nonumber\\
 && \!\!\!\!\!\!\!\!\!\!\!\!\!\!\!\!\!\!\!
    -m_{\ell^i} \bar \ell^i \ell^i -\frac{m_{\ell^i}}{v} \frac{\cos\alpha}{\cos\beta} H
    \bar \ell^i \ell^i +    \frac{m_{\ell^i}}{v} \frac{\sin\alpha}{\cos\beta} h
    \bar \ell^i \ell^i
%    \nonumber\\
%    &&
    -\frac{m_{\ell^i}}{v} z \bar \ell^i i \gamma_5 \ell^i +
      \frac{m_{\ell^i}}{v}\tan\beta A \bar \ell^i i\gamma_5 \ell^i.
 \end{eqnarray}

 The coupling constants are constrained by the conditions of theoretical
 consistencies such as perturbative unitarity~\cite{pu-thdm} or
 vacuum stability and triviality as a function of
 the cutoff scale $\Lambda$ of the model~\cite{rge-thdm,zee-ph}.
 We will take into account these conditions and work in the allowed parameter
 region for $\Lambda \gsim$ 10 TeV.

\section{Neutrino Mass and Mixing}

In this section we calculate the mass matrix in our model, which is
induced at the three loop level, and study the parameter regions which
satisfy the current data.
The mass term for LH neutrinos 
\begin{eqnarray}
  {\cal L}^{\rm eff} = \overline{\nu_L^c}^i M^\nu_{ij} {\nu_L}^j
 \end{eqnarray}
is naturally generated from dimension five operators
\begin{eqnarray}
   {\cal O}_5 = \frac{\xi_{ij}}{\Lambda_N}  \overline{{\nu}_L^c}^i
    {\nu_L}^j \phi \phi,
 \end{eqnarray}
in the low energy effective theory,  
where $\phi$ represents the neutral component of the Higgs doublet, 
$\xi_{ij}$ are dimensionless coefficients, and $\Lambda_N$ is a 
dimensionful scale of new dynamics. 
In ordinary models based on the seesaw mechanism, the tiny neutrino mass
is essentially realized by taking a very large scale for $\Lambda_N$~\cite{see-saw}:
$\Lambda_N=10^{13-15}$ GeV is required for $\xi_{ij}={\cal O}(1)$ to obtain
the tiny mass scale comparable to the neutrino data.
When $\Lambda_N$ is at most TeV scales, very small values for $\xi_{ij}$
are required to describe the data. It would be possible that such
small $\xi_{ij}$ would be  generated at loop level without making fine
tuning on the coupling constants in the Lagrangian.

\subsection{Evaluation of the three-loop induced neutrino mass matrix}

\begin{figure}[t]
\begin{center}
  \epsfig{file=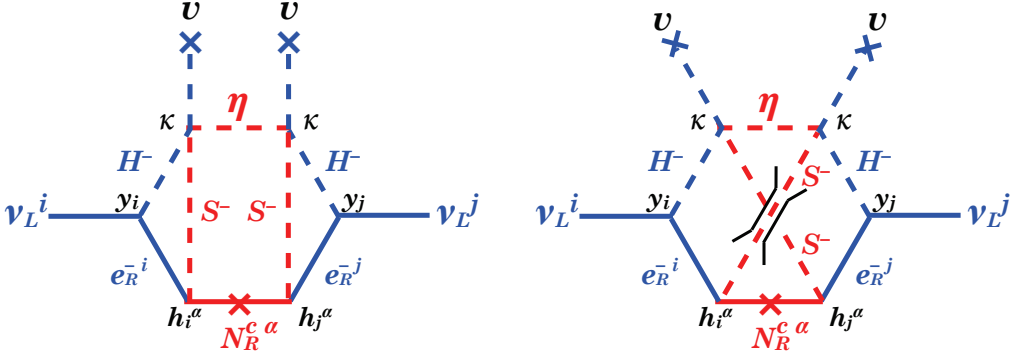,width=14cm}
\end{center}
  \caption{The diagrams for generating tiny neutrino masses. }
  \label{diag-numass}
\end{figure}
In our model, the LH neutrino mass matrix $M^\nu_{ij}$ is generated by the three-loop diagrams in
FIG.~\ref{diag-numass}.
The absence of lower order loop contributions is guaranteed by $Z_2$.  
The charged Higgs boson $H^\pm$ from the doublets and the charged
leptons $e_R^i$ play a crucial role to connect
LH neutrinos with the one-loop (box) diagram
by the $Z_2$-odd particles.
The resulting LH neutrino mass matrix is obtained as 
\begin{eqnarray}
M^\nu_{ij} &=& \sum_{\alpha=1}^2 
   C_{ij}^\alpha F(m_{H^\pm}^{},m_S^{},m_{N_R^{\alpha}}, m_\eta), \label{eq:mij}
\end{eqnarray}
where
\begin{eqnarray}
C_{ij}^\alpha &=&
   4 \kappa^2 \tan^2\beta 
  (y_{\ell_i}^{\rm SM} h_i^\alpha) (y_{\ell_j}^{\rm SM} h_j^\alpha), 
\end{eqnarray}
and the loop integral function $F$ is given by
\begin{eqnarray}
&&F(m_{H^\pm}^{},m_S^{},m_{N}, m_\eta) =
   \left(\frac{1}{16\pi^2}\right)^3 \frac{(-m_N^{})}{m_N^2-m_\eta^2}
 \frac{v^2}{m_{H^\pm}^4}\nonumber\\
 && \times \int_0^{\infty} x dx
  \left\{B_1(-x,m_{H^\pm}^{},m_S^{})-B_1(-x,0,m_S^{})\right\}^2
  \left(\frac{m_N^2}{x+m_N^2}-\frac{m_\eta^2}{x+m_\eta^2}\right),
\end{eqnarray}
with 
$m_f$ representing the mass of the field $f$ and $y_{e_i}^{\rm
SM}=\sqrt{2}m_{e_i}/v$. The function $B_1$ is the tensor coefficient
in the formalism by Passarino-Veltman 
for one-loop integrals~\cite{passarino-veltman}.
The detailed calculation of $F$ is shown in Appendix~B. %\ref{app:calcF}.

In Fig.~\ref{valueF}, we show the magnitude of the integral function $F$
as a function of $m_N^{}$ for several values of $m_{S^\pm}^{}$ and $m_\eta$.
It can be seen that $F$ becomes smaller for larger values of
$m_{S^\pm}^{}$. For $m_{S^\pm}^{} \lsim 500$ GeV $F$ decreases
 monotonically as $m_{N_R}^{}$ grows, and for greater values of
 $m_{S^\pm}^{}$ it mildly increases but finally turns to decrease
 as $m_{N_R}^{}$ grows. The dependences on $m_{N_R}^{}$ are however
 not very sensitive.
Numerically, the magnitude of $F$ is of order $10^{4}$ eV
in the wide range of parameter regions of our interest.  
We note that the dependence on the mass of the charged Higgs boson
$H^\pm$ is qualitatively similar to that on $m_{S^\pm}^{}$, so that $F$ becomes
smaller for larger values of $m_{H^\pm}^{}$.
\begin{figure}[t]
\begin{center}
  \epsfig{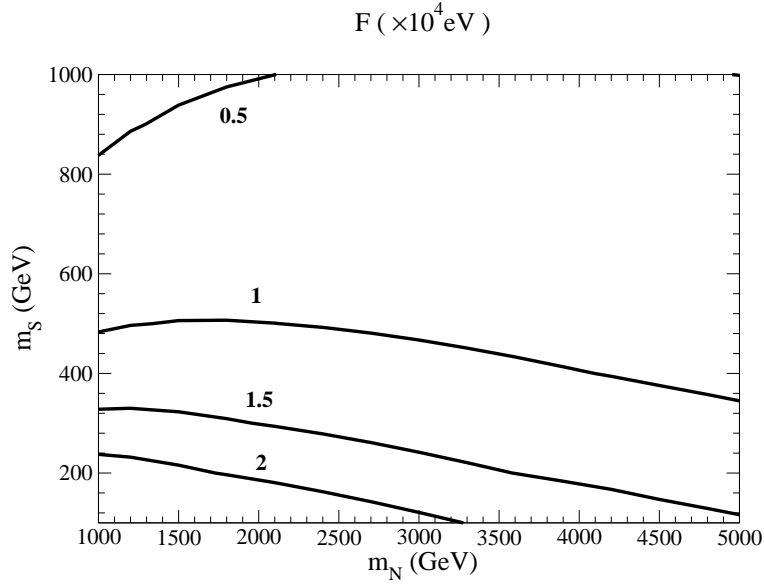}
\end{center}
  \caption{The contour plot of the values $F$ as a function of
 $m_{N}^{}$ and $m_{S}^{}$ for  $m_{\eta}=50$ GeV and $m_{H^\pm}^{}=100$
 GeV. }
  \label{valueF}
\end{figure}

Magnitudes of $h_e^\alpha$, $\kappa$ and $\tan\beta$
as well as $F(m_H^{},m_S^{},m_{N_R}, m_\eta)$ 
determine the universal scale of $M^\nu_{ij}$, 
whereas variation of $h_i^\alpha$ ($i=e$, $\mu$, $\tau$; $\alpha=$1-2)
reproduces the mixing pattern indicated by the neutrino
 data~\cite{lep-data}.

  \subsection{Parameters to reproduce the data of neutrino masses and mixings}

The generated mass matrix 
$M^\nu_{ij}$ in Eq.~(\ref{eq:mij}) of LH neutrinos
can be related to the neutrino oscillation data by
\begin{eqnarray}
 M^\nu_{ij} = U_{is} (M^\nu_{\rm diag})_{st} (U^T)_{tj}, 
\end{eqnarray}
where $M^\nu_{\rm diag}$ $=$ ${\rm diag}(m_1, m_2, m_3)$.
For the case of the normal hierarchy we identify the mass eigenvalues
as  
$m_1=0$, $m_2=\sqrt{\Delta m_{\rm solar}^2}$ and $m_3=\sqrt{\Delta m_{\rm
atm}^2}$, while  for inverted hierarchy $m_1=\sqrt{\Delta m_{\rm atm}^2}$,
$m_2=\sqrt{\Delta m_{\rm atm}^2 + \Delta m_{\rm solar}^2}$ and $m_3=0$
are taken.
The Maki-Nakagawa-Sakata matrix~\cite{mns} is parameterized as 
\begin{eqnarray}
  U=\left[\begin{array}{ccc}
     1&0&0\\
     0&c_{23}^{}&s_{23}^{}\\
     0&-s_{23}^{}&c_{23}^{}\\
           \end{array}
     \right]
  \left[\begin{array}{ccc}
     c_{13}^{}&0&s_{13}^{}e^{i\delta}\\
     0& 1 &0\\
     -s_{13}^{}e^{-i\delta}&0&c_{13}^{}\\
           \end{array}
     \right]
  \left[\begin{array}{ccc}
     c_{12}^{}&s_{12}^{}&0\\
     -s_{12}^{}&c_{12}^{}&0\\
         0&0&1
          \end{array}
     \right]
    \left[\begin{array}{ccc}
     1&0&0\\
     0&e^{i\tilde{\alpha}}&0\\
         0&0&e^{i\tilde{\beta}}
          \end{array}
     \right],
\end{eqnarray}
where $s_{ij}$ and $c_{ij}$ represent $\sin\theta_{ij}$ and
$\cos\theta_{ij}$ respectively with $\theta_{ij}$ to be the neutrino mixing angle between the
$i$th and $j$th generations, and $\delta$ is the Dirac phase while
$\tilde{\alpha}$ and $\tilde{\beta}$ are Majorana phases.
For simplicity, we neglect the effects of these phases in the following
analysis.
Current neutrino oscillation data give the following 
values~\cite{lep-data};  
\begin{eqnarray}
\Delta m^2_{\rm solar} \simeq 7.65\times 10^{-5} \,{\rm eV}^2 \,,~~
|\Delta m^2_{\rm atm}| \simeq 2.4\times 10^{-3} \,{\rm eV}^2\,, \\
\sin^2\theta_{12} \simeq 0.3 \,,~~~~
\sin^2\theta_{23} \simeq 0.5 \,,~~~~
\sin^2\theta_{13} < 0.04\,.~~~~~~
\label{obs_para}
\end{eqnarray}

In order to study parameter sets that satisfy the current experimental
data, we need to discuss constraints from the $\mu \to e \gamma$ results.
This process is induced by one loop diagram of a right handed neutrino $N_R$ 
and a charged scalar boson $S^\pm$ through Yukawa coupling $h_i^\alpha$
($i=e$ and $\mu$). The branching ratio is given by
\begin{equation}
B(\mu\to e \gamma) \simeq  \frac{3\alpha_{\rm em} v^4}{32\pi}\left[
\frac{|h_e^1 h_\mu^1|^2}{m_{S^\pm}^4}
\left\{F_2\left(\frac{m_{N_R^1}^2}
{m_{S^\pm}^2}\right)\right\}^2+
\frac{|h_e^2 h_\mu^2|^2}{m_{S^\pm}^4}\left\{F_2\left(\frac{m_{N_R^2}^2}
{m_{S^\pm}^2}\right)\right\}^2
\right] ,
\end{equation}
where $F_2(x)\equiv (1-6x+3x^2+2x^3-6x^2-\ln x)/6(1-x)^4$.
Comparing this to the current experimental bound, $B(\mu\to e \gamma)
< 1.2 \times 10^{-11} $~\cite{MEGA}, 
the mass of $N_R$ or $S^\pm$ is strongly constrained from below.

Under the {\it natural} requirement that  
$h_e^\alpha = {\cal O}(1)$, and taking into account 
the $\mu\to e\gamma$ search results~\cite{MEGA},   
we find that $m_{N_R^\alpha}^{} \gsim {\cal O}(1)$ TeV,
$m_{H^\pm}^{} = 100$-200 GeV and $\kappa \tan\beta = {\cal O}(10)$. 
In addition, with the LEP direct search data for Higgs bosons and precision measurement
data~\cite{lep-data}, possible values for the scalar masses uniquely turn out to be  
$m_{H^\pm}^{} \simeq m_{H}^{} \simeq 100$ GeV and $m_{S^\pm}^{} \sim
{\cal O}(100)$ GeV for $\sin(\beta-\alpha) = 1$.
As we see in later sections, relatively heavier $S^\pm$ ($m_{S^\pm}^{}
\gsim 400$ GeV) is favored from the 
discussion on DM and electroweak baryogenesis.
It is known that in the Type X Yukawa interaction a light $H^\pm$
($m_{H^\pm}^{} \lsim 300$ GeV) is not excluded by the $b \to s \gamma$ data~\cite{bsgamma}.
This is the reason that the Yukawa coupling in Eq.~(\ref{eq:yukawa1})
has been employed in our model.

\begin{table}
\begin{center}
%\underline {{\bf STANDARD ($m_\eta$=50 GeV): } } \\[2mm]
  \begin{tabular}{|l||c|c|c|c||c|c|c|c|c|c||c|}\hline
%\multicolumn{2}{|c||}
{\mbox{Set}}  & \multicolumn{3}{c|}{\mbox{Mass (TeV)}} & &
   \multicolumn{6}{c||}{\mbox{Coupling~constants}} & LFV\\ \hline
      ~(hierarchy,~$\sin^22\theta_{13}$)& $m_\eta$ & $m_S$ &$m_{N_R^\alpha}^{}$&$\kappa\tan\beta$ & $h_e^1$ & $h_e^2$ & $h_\mu^1$ & $h_\mu^2$ & $h_\tau^1$ & $h_\tau^2$  &
   $B(\mu\!\!\to\!\! e\gamma)$\\\hline \hline
   A~(normal, 0) &0.05 & 0.4 &3 &  29 &  2.0 &  2.0 &   0.041  &-0.020 &  0.0012 &  -0.0025
   &$6.8\!\times \!10^{-12}$ \\ \hline 
B~(normal, 0.14)&  0.05 & 0.4 &3 & 34 & 2.2 &  2.1  & 0.0087  & 0.037 & -0.0010 &  0.0021
  &$5.3\!\times \!10^{-12}$ \\ \hline 
      \hline 
        C~(inverted, 0) & 0.05 &0.4 & 3 &  66 & 3.8 &  3.7  & 0.013  &-0.013 & -0.00080 &  0.00080
                                       & $4.2\!\times \!10^{-12}$ \\ \hline 
        D~(inverted, 0.14) &0.05 & 0.4 & 3 &  66 &  3.7 &  3.7 & -0.016  & 0.011 &  0.00064 & -0.00096 
                   &$4.2\!\times \!10^{-12}$ \\ \hline 
   \end{tabular}
\end{center}
\vspace{-2mm}  \caption{Values of $h_i^\alpha$ which satisfy neutrino
 data and the constraint from $\mu \to e \gamma$
 for $m_\eta=50$ GeV, $m_{H^\pm}^{}=100$GeV and $m_{N_R^1}=m_{N_R^2} $. }
  \label{h-numass}
 \end{table}
For values of $h_i^\alpha$ ($(i=\mu, \tau)$), we only require that $h_i^\alpha y_i \sim {\cal O}(y_e) \sim
10^{-5}$, since we cannot avoid to include the hierarchy among $y_i^{\rm SM}$. 
Several sets for $h_i^\alpha$ which satisfy the neutrino data are shown
in TABLE~\ref{h-numass} with the predictions on the branching ratio
of $\mu\to e\gamma$ for $m_\eta=50$ GeV, $m_{H^\pm}=100$ GeV and $m_{N_R^1} = m_{N_R^2} = 3$ TeV. 
Set A and Set B are rather standard choices in our model assuming the
normal hierarchy, while C and D are those assuming the inverted hierarchy.
We note that larger values are required for $\kappa \tan\beta$ in 
the case of the inverted hierarchy. 
On the other hand, small $\tan\beta$ is favored by 
the realization of electroweak baryogenesis as we will discuss
later\footnote{
In the THDM, $\tan\beta \gsim 10$ would not be favored for successful
electroweak baryogenesis under the EDM constraints~\cite{ewbg-thdm3}.
Although we do not discuss CP violating effects in our present analysis,
we keep in mind this constraint and just impose the condition that
$\tan\beta$ is not larger than about 10.}. 
Therefore, our model turns out to be better compatible with the normal hierarchy scenario~\cite{comment1}.

Set E and Set F are chosen to potentially include the results from
PAMELA (and ATIC) and those from DAMA, respectively, as discussed
later. The parameters describing the data are listed in Table~\ref{h-numass2}. 
\begin{table}
\begin{center}
%\underline {{\bf PAMELA ($m_\eta$=700 GeV) and DAMA ($m_\eta$=5 GeV): } } \\[2mm]
  \begin{tabular}{|l||c|c|c|c||c|c|c|c|c|c||c|}\hline
%\multicolumn{2}{|c||}
{\mbox{Set}}  & \multicolumn{3}{c|}{\mbox{Mass (TeV)}} & &  \multicolumn{6}{c||}{\mbox{Yukawa~couplings}} & LFV\\ \hline
      ~~(hierarchy,~$\sin^22\theta_{13}$)& $m_\eta$ & $m_S$ &$m_{N_R^\alpha}^{}$&$\kappa\tan\beta$ & $h_e^1$ & $h_e^2$ & $h_\mu^1$ & $h_\mu^2$ & $h_\tau^1$ & $h_\tau^2$  &
   $B(\mu\!\!\to\!\! e\gamma)$\\\hline \hline
         E~(normal,  0) &0.7 &     0.9&3 & 33 &  3.0  &2.0 & -0.014   &0.057 & -0.0028 &  0.0021
                  &   7.2 $\!\times \!10^{-12}$ \\ \hline 
\hline
         F~(normal,  0) &0.005 &  0.4 &3 &29 &  
          2.0   &2.0 & -0.019  & 0.041  &-0.0024 &  0.0012
          & 6.6  $\!\times \!10^{-12}$ \\ \hline 
   \end{tabular}
\end{center}
\vspace{-2mm}  \caption{Values of $h_i^\alpha$ which satisfy neutrino
 data and the constraint from $\mu \to e \gamma$ for
 $m_\eta=700$ GeV (Set E) and $m_\eta=5$ GeV (Set F).
 The other parameters are taken to be $m_{H^\pm}^{}=100$GeV and
 $m_{N_R^1}=m_{N_R^2} $ for the normal hierarchy.
 }
  \label{h-numass2}
 \end{table}

We also examined the case with three generations for RH neutrinos.
Unlike the case with two generations, the mass of the smallest
mass eigenvalue for LH neutrinos can be nonzero.
It is however found that larger values for $\kappa\tan\beta$
are prefered to reproduce the neutrino data, so that
it would be rather unnatural as compared to the case with two
generations.
Therefore, we concentrate on the model with two generation RH neutrinos 
in the rest of the paper.
 
\begin{figure}[tb]
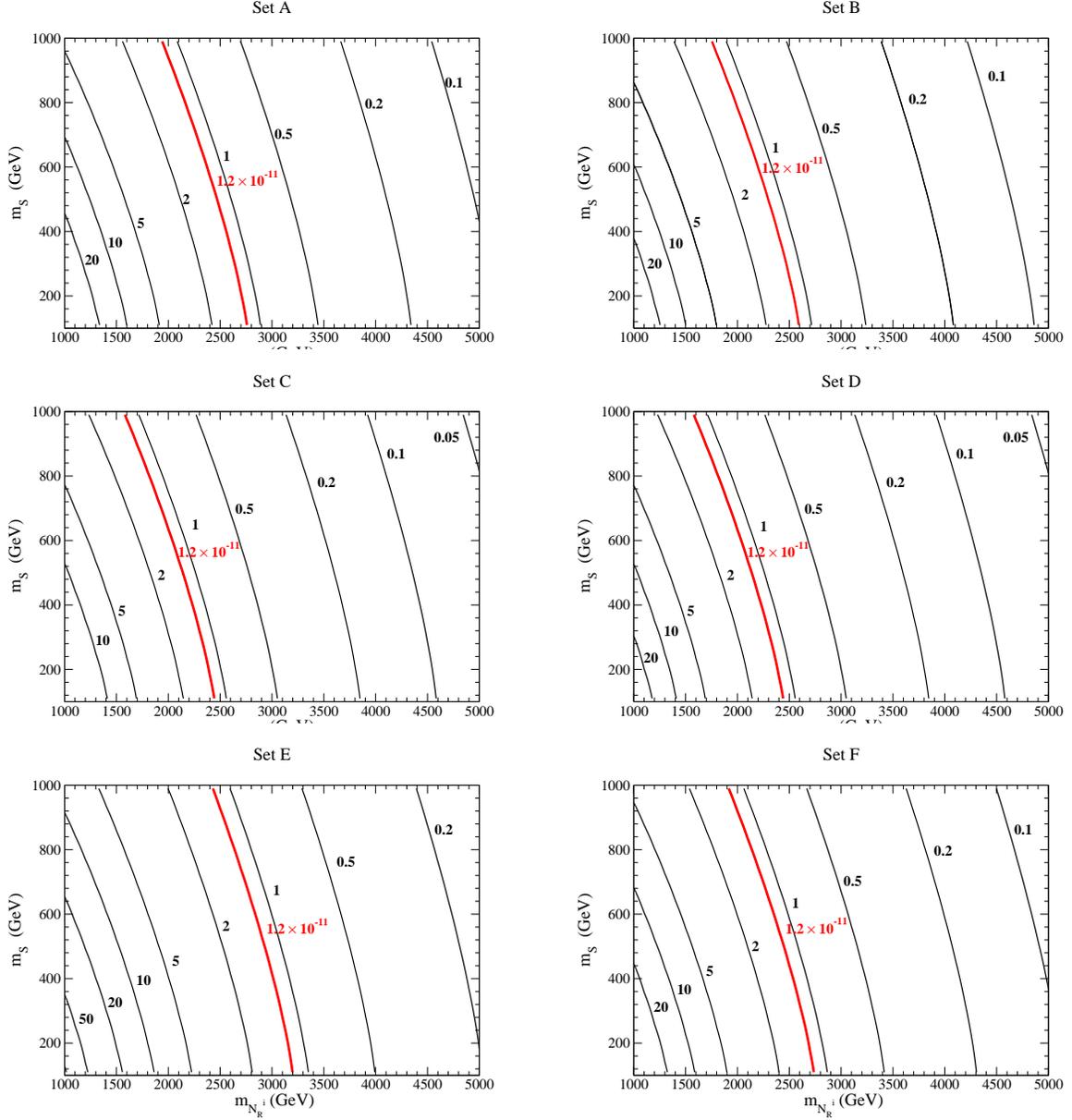

\begin{minipage}{0.49\hsize}
\includegraphics[width=7cm]{LFV_A.eps}
\end{minipage}
\begin{minipage}{0.49\hsize}
\includegraphics[width=7cm]{LFV_B.eps}
\end{minipage}\\
\begin{minipage}{0.49\hsize}
\includegraphics[width=7cm]{LFV_C.eps}
\end{minipage}
\begin{minipage}{0.49\hsize}
\includegraphics[width=7cm]{LFV_D.eps}
\end{minipage}\\
 \begin{minipage}{0.49\hsize}
\includegraphics[width=7cm]{LFV_E.eps}
 \end{minipage}
\begin{minipage}{0.49\hsize}
\includegraphics[width=7cm]{LFV_F.eps}
\end{minipage}
  \caption{The contour plot of $\mu\to e\gamma$ branching ratio in the
 $m_{N}^{}-m_S$ plane for the values of $h_i^\alpha$ given in Set A to  Set F. }
  \label{valueBr2_all}
\end{figure}

In Fig.~\ref{valueBr2_all}, the contour plots for the branching ratio
of $\mu\to e\gamma$ as a function of $m_{N_R}^{}$ and
$m_{S^\pm}^{}$ for values of $h_i^\alpha$ to be of Set A - Set F in
Table II and III at $m_{H^\pm}^{}=100$ GeV.
The dependence of the branching ratio on the mass of $S^\pm$ is rather mild.
On the other hand, the branching ratio is strongly dependent on the mass of $N_R$, so that the lower bound on $m_{N_R}^{}$ is obtained for a mass of $S^{\pm}$.

\section{Dark Matter}
\indent

Since $Z_2$ is exact, the lightest $Z_2$-odd particle is 
stable and can be a candidate of DM if it is neutral.
In our model, $N_R^\alpha$ must be heavy ($m_{N_R^\alpha}^{} \sim $ a
few TeV), so that 
the DM candidate is identified as $\eta$.
Since $\eta$ is a singlet under the SM gauge group,
the interactions with $Z_2$ even particles are only through the Higgs coupling.
It is, however, worth noticing the presence of interactions through $\kappa$ coupling, 
which is absent in the simplest gauge singlet scalar DM~\cite{john}. 

\subsection{Thermal relic abundance}

\begin{figure}[t]
\begin{center} 
  \epsfig{file=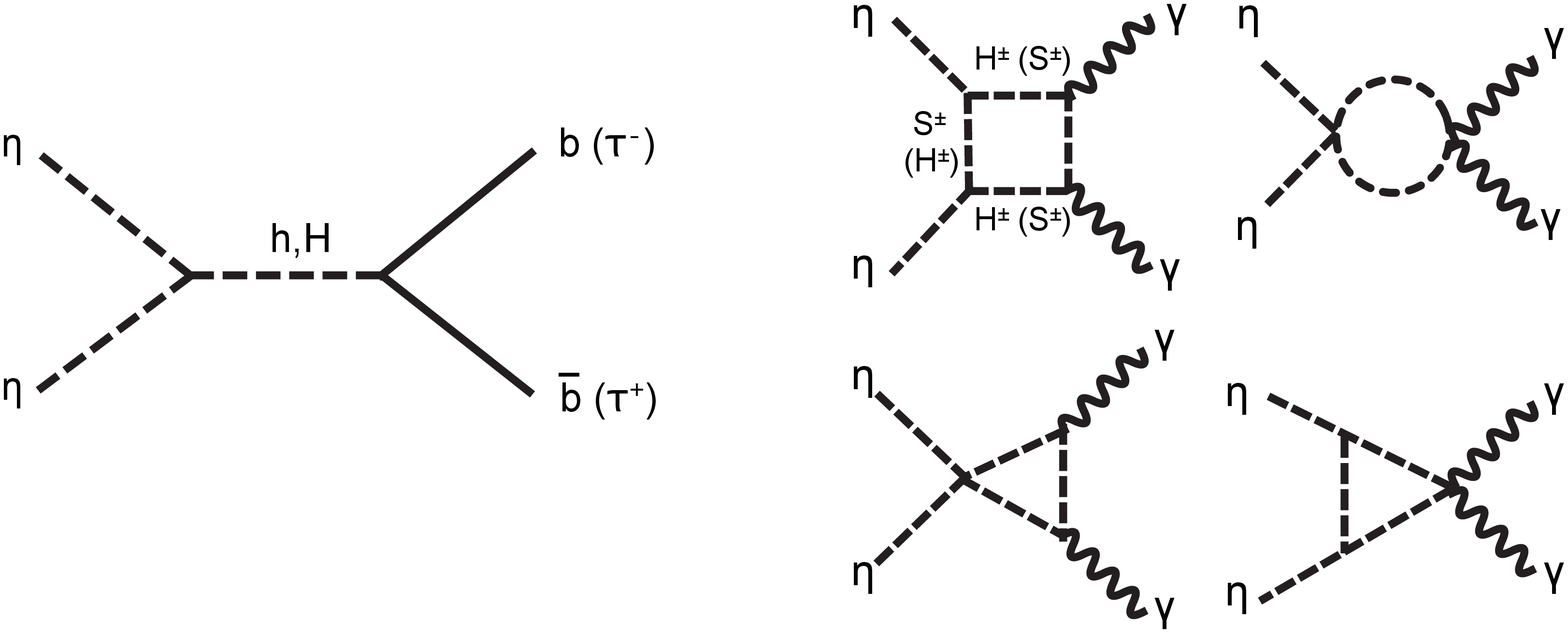,width=12.cm}
\end{center}
 \vspace{-2mm} 
  \caption{Feynman diagrams for the DM annihilation. For the one-loop
 diagrams for $\eta\eta\to\gamma\gamma$ only those which contain the
 coupling constant $\kappa$ are shown because they give dominant
 contributions in our parameter choices. 
 }
  \label{fig:AKS_DM_diag}
\end{figure}

In principle, $\eta$ has various annihilation processes 
into $\gamma\gamma, f \bar{f}, W^+ W^-$ and so on. 
The annihilation into a $W$ boson pair or a Higgs pair leads to 
too rapid annihilation.
Thus, relatively light $\eta$ would be favored 
in order to kinematically close these annihilation modes.
When $\eta$ is lighter than the $W$ boson, $\eta$ predominantly annihilates 
into two photons through one-loop diagrams by $H^\pm$ and $S^\pm$, 
as well as  into $b \bar{b}$ and $\tau^+\tau^-$ through $s$-channel
Higgs ($h$ and $H$) exchange diagrams: see Fig.~\ref{fig:AKS_DM_diag}.

The relevant annihilation rate is evaluated for
 $\sin(\beta-\alpha)\simeq 1$ as  
\begin{eqnarray}
&&\!\!\!\!\!\!\! w(s) \!
\simeq\!
 \frac{1}{16\pi}\left(\frac{e^2}{16\pi^2}\right)^2(\kappa v)^4
 \left(\frac{8}{m_S^2}\right)^2
 + \frac{3s m_b^2 v^2}{16\pi}\!
  \left| \frac{\sigma_1 \cos^2\beta +\sigma_2 \sin^2\beta}{s-m_h^2+i m_h
   \Gamma_h}\!  - \! \frac{(\sigma_1+\sigma_2)\cos^2\beta}
 {s-m_H^2+i m_H \Gamma_H} \right|^2  \nonumber \\
 &&\!\!\!\!\!\!\! + \frac{s m_\tau^2 v^2}{16\pi}\! 
 \left| \frac{\sigma_1\cos^2\beta+\sigma_2 \sin^2\beta}
 {s-m_h^2+i m_h \Gamma_h}   \!+\!  \frac{(\sigma_1-\sigma_2)  \sin^2\beta}
 {s-m_H^2+i m_H \Gamma_H} \right|^2, 
 \label{annihiration}
\end{eqnarray}
where $\Gamma_h$ and $\Gamma_H$ are the decay widths of $h$ and $H$ respectively, 
and $s$ is a usual Mandelstam variable.
The second and third term correspond to the annihilation into fermion 
and anti-fermion. 
The annihilation into two gammas is given by the first term whose 
expression was obtained under the s-wave approximation.
From their thermal averaged annihilation rate $\langle \sigma v \rangle$,
the relic mass density $\Omega_\eta h^2$ is evaluated 
 for $\eta\eta \to \gamma\gamma, b\bar b$ and $\tau^+\tau^-$ 
as   
\begin{eqnarray}
 \Omega_\eta h^2= 1.1\times 10^9
  \left.\frac{(m_{\eta}/T_d)}{ \sqrt{g_\ast} M_P \langle \sigma
   v\rangle } \right|_{T_d} \hspace{2mm} {\rm GeV}^{-1}, 
 \end{eqnarray}
where $M_P$ is the Planck scale, $g_\ast$ is the total number of
relativistic degrees of freedom in the thermal bath, and $T_d$ is the decoupling temperature~\cite{kt}.

Except for the region near the resonance annihilation by s-channel Higgs ($h$ and $H$) exchange
where $m_{\eta} \simeq m_h/2$ and $m_H/2$,
the DM relic density is basically determined by the annihilation into $\gamma\gamma$
whose cross section essentially depends on only two free parameters; $\kappa$ and $m_{S^\pm}^{}$.
Since we already know $\kappa = {\cal O}(1)$ from neutrino masses discussed 
in the previous section, we obtain $m_{S^\pm}^{} \gsim 400$ GeV from the $\eta$ DM abundance result.

\begin{figure}[t]
\begin{minipage}{0.49\hsize}
\includegraphics[width=8cm]{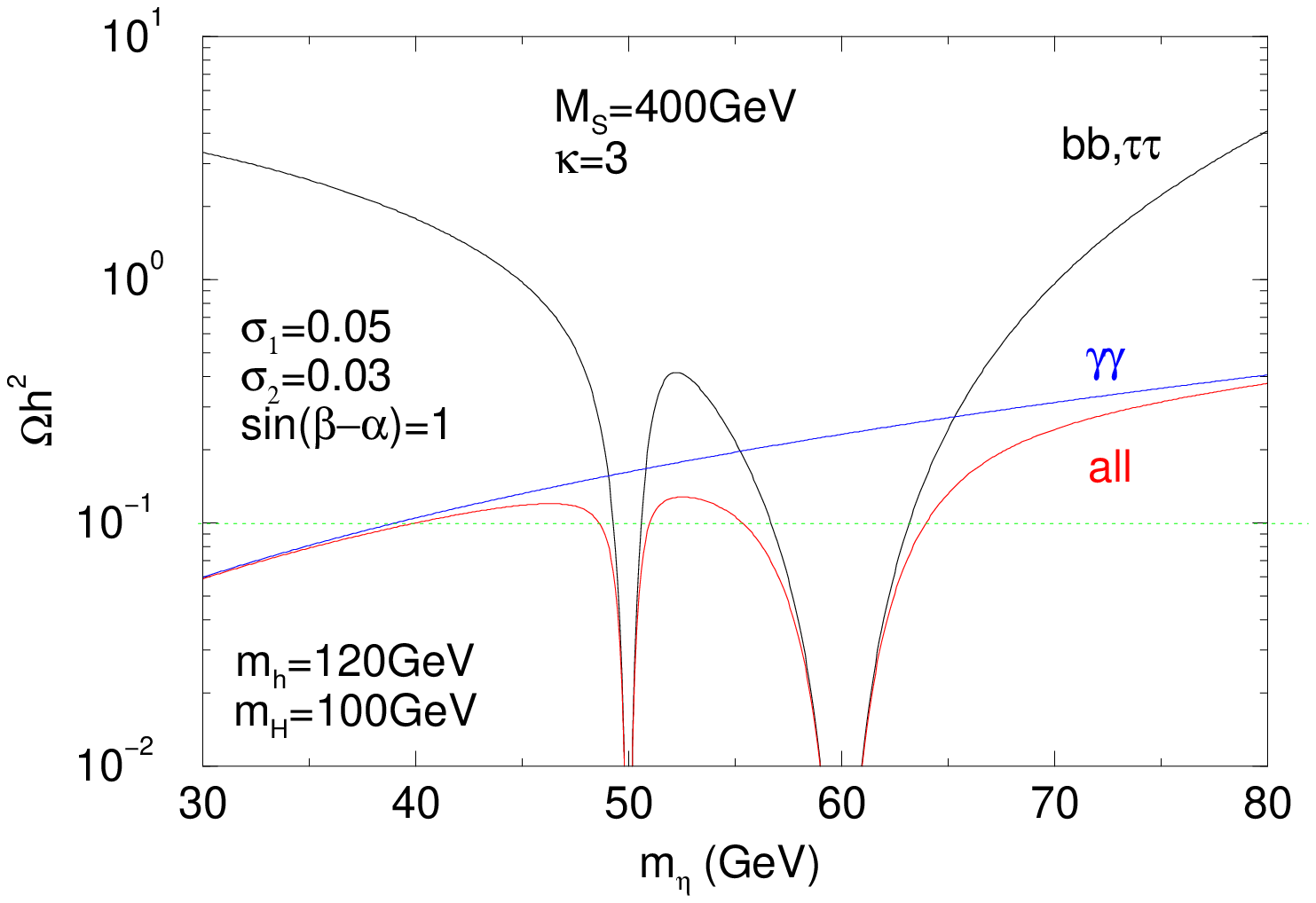}
\end{minipage}
\begin{minipage}{0.49\hsize}
\includegraphics[width=8cm]{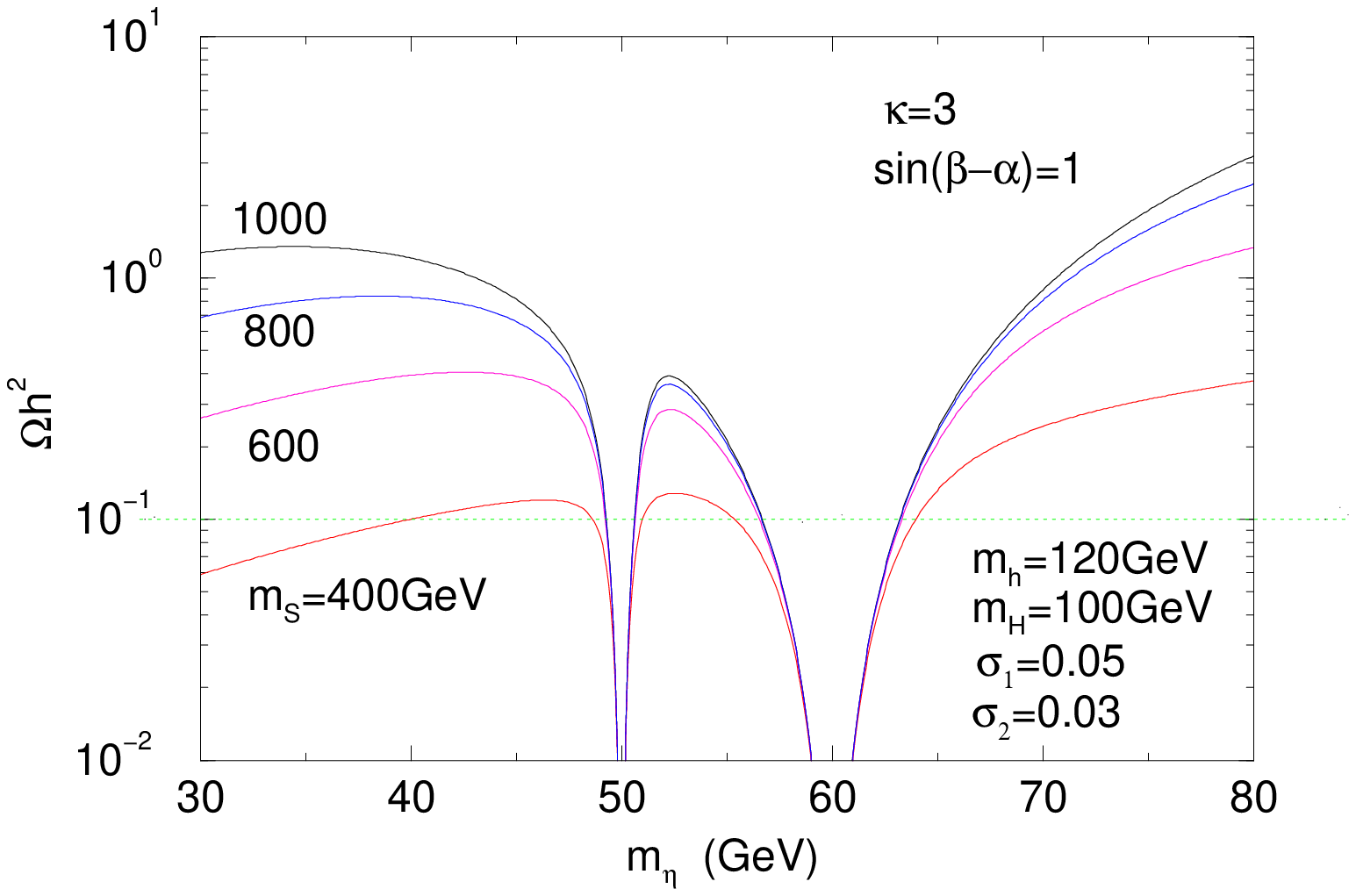}
\end{minipage}\\
 \vspace{-2mm}
  \caption{
 [Left] The thermal relic abundance of $\eta$ and contributions
 from tree and one-loop processes.
[Right] The thermal relic abundance of $\eta$ and its $m_{S^{\pm}}$ dependence.
 In both figures, parameters are taken as $(\kappa,\sigma_1, \sigma_2, m_h, m_H^{},\tan\beta) = (3,0.05, 0.03, 120
 {\rm GeV}, 100 {\rm GeV},10)$.
 }
  \label{etaOmega}
\end{figure}
FIG.~\ref{etaOmega} shows
$\Omega_{\eta}h^2$ as the function of $m_\eta$ for $\sin(\beta-\alpha)=1$.
The parameters are chosen as
$(\sigma_1, \sigma_2, \kappa\tan\beta, m_h, m_H)
 = (0.05, 0.03, 30, 120 {\rm GeV}, 100 {\rm GeV})$.
Strong annihilation can be seen near $50$ GeV $\simeq m_H^{}/2$
($60$ GeV $\simeq m_h/2$) due to the resonance of $H$ ($h$) mediation.
The $\Omega_{\eta}h^2$ is proportional to $m_{\eta}^2$ for the dominant annihilation into $\gamma\gamma$.
The data ($\Omega_{\rm DM} h^2 \sim 0.11$~\cite{wimp}) indicate that $m_\eta$ is around $50-65$ GeV 
for $m_{S^\pm}^{}$ of $400$ GeV.
If $S$ is much lighter than $400$ GeV, 
the resultant $\Omega_{\eta} h^2$ is below $0.1$ for most of $m_{\eta}$ range. 
For a heavier $m_{S^\pm}^{}$, only the resonance annihilation would provide the desired $\Omega_{\eta}h^2$. \\

\subsection{Constraints from DM direct searches}

Generally speaking, WIMP DM is detectable 
by direct DM searches  such as CDMS~\cite{CDMS}.
Scatterings off of WIMP DM with nuclei take place in two different ways, 
depending on the nature of WIMP~\cite{Jungman:1995df}.
One is spin-dependent scattering whose cross section depends on 
the total spin of target nuclei, while the other is spin-independent(SI).
$\eta$ is a scalar particle, therefore 
the relevant process is a SI scattering 
through t-channel Higgs ($h$ and $H$) exchange~\cite{john}.
The SI cross section for a proton is 
given as~\cite{Jungman:1995df}
\begin{eqnarray}
\sigma_p^{SI}
 =  \frac{m_p^2}{\pi (m_{\eta}+m_p)^2}f_p^2 ,  
\end{eqnarray}
with
\begin{eqnarray}
 \frac{f_p}{m_p} =
  \sum_{q=u,d,s} f_{T\,q}^{(p)}\frac{f_q}{m_q}
   + \frac{2}{27}\sum_{q=c.b,t} f_{TG\,q}^{(p)}\frac{f_q}{m_q},
\end{eqnarray}
where $m_p$ and $m_q$ are the proton and each flavor quark masses. 
The hadronic matrix elements $f_{T q}^{(p)}$ and $f_{TG q}^{(p)}$ are defined as
\begin{eqnarray}
 m_p f_{T\,q}^{(p)} \equiv \langle p|m_q \bar{q}q|p\rangle , {\hspace{6mm}}
 f_{TG\,q}^{(p)} = 1-\sum f_{T\,q}^{(p)},
\end{eqnarray}
and we adopted the values in Ref.~\cite{Ellis:2008hf}
 for the following estimation.
In our model, the coupling constant of the effective interaction
 between $\eta$ and quark, ${\cal L}_{int} \supset f_q \bar{q}q\eta\eta$, 
 is given as
\begin{eqnarray}
 \frac{f_q}{m_q} = \frac{(-\sigma_1\sin\alpha\cos\beta+\sigma_2\cos\alpha\sin\beta)}{2m_h^2}
\frac{\cos\alpha}{\sin\beta} 
+\frac{(\sigma_1\cos\alpha\cos\beta+\sigma_2\sin\alpha\sin\beta)}{2m_H^2}\frac{\sin\alpha}{\sin\beta}.
\end{eqnarray}
Until now, only null results are reported from 
all experiments of direct DM search.
The current most stringent bounds are given by XENON 10~\cite{xenon} 
and CDMS II as $\sigma_p^{SI} \lesssim 5 \times 10^{-8}$ pb.
We find the coupling constants $\sigma_i \lesssim {\cal O}(10^{-2})$ are consistent.

Remember the most of parameter region of 
the minimal singlet scalar DM is excluded by null results of DM direct search 
and only resonance annihilation region, 
which would mean certain fine-tuning between DM mass and Higgs mass, 
is consistent~\cite{john}.
Hence, in contrast with this, it is worth emphasizing that 
such a tuning to realize large resonance is not necessarily required 
in our $\eta$ DM, because the one-loop processes give 
sizable contributions to annihilation for $m_S \simeq 400$ GeV. 
Another singlet scalar DM model without necessity of a resonance 
due to different mechanism is also recently proposed~\cite{Cerdeno}.

\subsection{Variations with a particular choice of parameters}

So far, we have discussed properties of $\eta$ DM 
on a standard basis according to parameter sets in Table~\ref{h-numass}.
However, there are a few recent observational results which would imply that 
the nature of DM could differ from the standard WIMP.

The first example is anomalous excesses in the observed positron flux.
The PAMELA data shows that observed positron flux exceeds 
the normally expected astrophysical background flux 
for energy range above a few hundred GeV 
with a rising spectrum 
in $\Phi^{e^+}/(\Phi^{e^-}+\Phi^{e^+})$~\cite{pamela}.
In addition, 
ATIC also reported a bump around $E \sim 600$ GeV in the spectrum~\cite{atic}.
These excesses could be interpreted as a signal from 
annihilation~\cite{annihilation} or decay~\cite{decay} of DM particles.
As mentioned above, we have argued that the mass of $\eta$ 
is most likely around $50$ GeV in order to avoid its annihilation 
into $W$-boson or Higgs boson.
However, if we accept negligibly tiny $\sigma_i$ coupling constants 
to suppress these annihilation via the s-channel Higgs exchange, 
it is possible to realize that $\eta$ is as heavy as several hundreds
GeV in Set E in Table~\ref{h-numass2}. 
Indeed, in such a case, 
the annihilation into a pair of charged Higgs bosons, 
which subsequently decays dominantly into a tau lepton and a neutrino 
due to the Type-X Yukawa interaction,
becomes the dominant mode and the resultant $\Omega_{\eta}h^2 \simeq 0.1$ 
corresponds to $\eta$ mass of several hundreds GeV~\footnote{
 We estimated this from Eq.~(\ref{annihiration}), 
 which might be too simple for a heavy $\eta$. 
 However, we suppose that the error would be just a factor difference.
 }.
The tau lepton produces electrons and positrons as well as hadrons.
Hence, at least as far as the energy scale is concerned, 
the ATIC anomaly might be account for a heavy $\eta$.
The detailed calculation, paying attentions to the spectrum,  
will be presented elsewhere~\cite{AKSpamela}\footnote{Within
the framework of three loop generation of neutrino masses,
the explanation of positron excess was already
examined in Ref.~\cite{kingman} by extending the other model
in Refs.~\cite{knt,kingman-seto}.}.

\begin{figure}[t]
\begin{center} 
  \epsfig{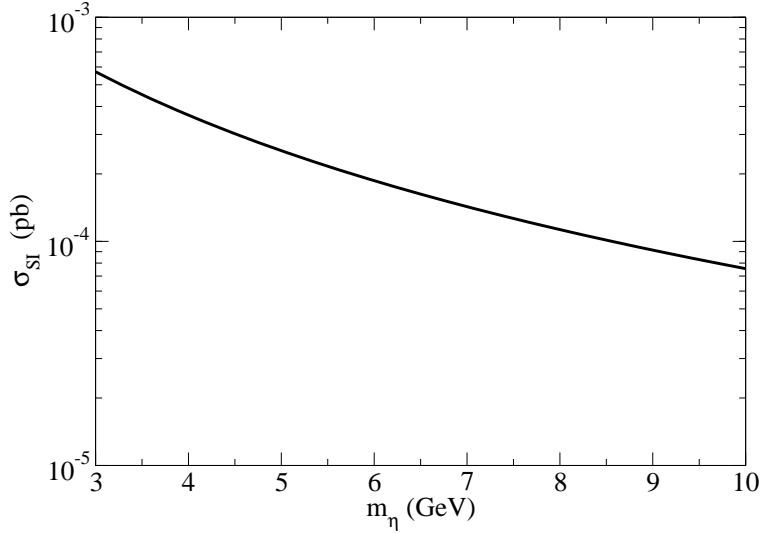}
\end{center}
 \vspace{-2mm} 
  \caption{A large enough scattering cross section of light DM with a proton 
  to account for the DAMA$/$LIBRA claim.
 }
  \label{damaSI}
\end{figure}

Next, the results at the DAMA$/$LIBRA experiment
claim a significant annual modulation in their DM detection 
rate which also could be caused by WIMPs~\cite{dama}, 
whereas all other experiments do not find any signal yet.
These two results appear to conflict each other, 
but a few compatible scenarios are possible, e.g., 
inelastic scattering DM~\cite{inelastic}, nevertheless.
One of them is a light WIMP with 
the mass of $3 \,{\rm GeV}\lesssim m_{\eta} \lesssim 8 \,{\rm GeV}$  and 
a large scattering cross section with nuclei 
of ${\cal O}(10^{-4})$ pb~\cite{lightWIMP} (as
in Set F in Table~\ref{h-numass2}).
Such a large $\sigma_p^{SI}$ can be available for $\sigma_2 \sim 0.4$ as seen in FIG.~\ref{damaSI}.
With a slightly heavier charged scalar $S$ to suppress one-loop contributions,
 the desired thermal abundance $\Omega_{\eta}h^2 \simeq 0.1$ can be obtained
 almost independently from $m_{\eta}$ in the relevant mass range.

\section{Strong First Order Phase Transition for Successful Electroweak Baryogenesis}

\indent
The model satisfies the Sakharov's conditions for
baryogenesis~\cite{sakharov}:  
baryon number violation occurs via sphaleron effect at high temperatures, C and CP
violation is automatic in the electroweak theory and additional CP
violating phases are in the Higgs sector and in the Yukawa interaction, and 
the condition of departure from thermal equilibrium can be realized by
the strong first order EWPT in the electroweak baryogenesis scenario.
In this section, we examine the region in the parameter space
where the strong first order EWPT is realized. 

In this paper, we take the SM-like limit for the mixing between
$h$ and $H$ ($\sin(\alpha-\beta)=-1$), where
$h$ is the SM-like Higgs boson while $H$ does not have VEV. 
In such a case, the one-loop (both zero and finite temperature)
effective potential
is described in terms of
a unique order parameter $\varphi$ ($=\langle h\rangle$)\footnote{The
effect of the two stage phase transition in the case with a
large $\tan\beta \gg 1$ is neglected in our analysis~\cite{twostage}.}.
All extra scalar bosons ($H$, $A$, $H^\pm$, $S^\pm$, $\eta$)
approximately behave just like additional particles running in the loop
in the effective potential.
The following relations and equations do not depend on the value of
$\tan\beta$, so that we apply our result here to the case of $\tan\beta
= {\cal O}(1)-{\cal O}(10)$.

\subsection{Effective potential}

We first consider the one-loop effective potential at zero temperature
$V_{eff}[\varphi] = V_{\rm tree}[\varphi] + \Delta V[\varphi]$. 
The one-loop contribution $\Delta V[\varphi]$ is given by 
\begin{eqnarray}
   \Delta V[\varphi] = \frac{1}{64\pi^2} \sum_f 
                     N_{c_f} N_{s_f} (-1)^{2 s_f}  
                     (M_f[\varphi])^4 
             \left\{ \ln \frac{(M_f[\varphi])^2}{Q^2}
                         -\frac{3}{2}\right\},  
\label{eff}
\end{eqnarray}
where $\varphi=\langle \phi \rangle = v + \langle h \rangle$, 
$N_{c_f}$ is the color factor,   
$s_f$ ($N_{s_f}$) is the spin (degree of freedom) of the field $f$  
in the loop, $M_f[\varphi]$ is the field dependent mass of $f$, 
and $Q$ is an arbitrary scale.  For $\sin(\beta-\alpha)=1$, 
the field dependent mass functions are given by 
\begin{eqnarray}
  {\tilde m}_h^2(\varphi) &=& \frac{3 m_h^2}{2} \left(\frac{\varphi^2}{v^2} -\frac{1}{3}\right), \\ 
  {\tilde m}_H^{2}(\varphi) &=& 
  \left(m_H^2 - M^2+\frac{m_h^2}{2}\right) \frac{\varphi^2}{v^2}+M^2-\frac{m_h^2}{2}, \\ 
  {\tilde m}_{H^\pm}^{2}(\varphi) &=&  
  \left(m_{H^\pm}^2 - M^2+\frac{m_h^2}{2}\right)\frac{\varphi^2}{v^2}+M^2-\frac{m_h^2}{2}, \\ 
  {\tilde m}_A^2(\varphi) &=& 
\left(m_A^2 - M^2+\frac{m_h^2}{2}\right)
\frac{\varphi^2}{v^2}+M^2-\frac{m_h^2}{2}, \\
  {\tilde m}_S^2(\varphi) &=& 
\left(m_S^2 - \mu_S^2\right)
\frac{\varphi^2}{v^2}+\mu_S^2, \\
  {\tilde m}_\eta^2(\varphi) &=& 
\left(m_\eta^2 - \mu_\eta^2\right)
\frac{\varphi^2}{v^2}+\mu_\eta^2. 
\end{eqnarray}
The renormalized mass of the SM like Higgs boson $h$ and the one-loop corrected $hhh$ coupling
are given by the conditions 
\begin{eqnarray}
\left.
\frac{\partial}{\partial \varphi} V_{eff}[\varphi] \right|_{\varphi=v} &=& 0,\\
\left.
\frac{\partial^2}{\partial \varphi^2} V_{eff}[\varphi] 
\right|_{\varphi=v} &=& m_h^2, \\
\left.
\frac{\partial^3}{\partial \varphi^3} V_{eff}[\varphi] 
\right|_{\varphi=v} &=& \lambda_{hhh}^{\rm eff},
\end{eqnarray}
when $\sin(\beta-\alpha)=1$.

\subsection{Electroweak phase transition}

The additional part of the one-loop effective potential 
at finite temperature is expressed by  
\begin{eqnarray}
   \Delta V_T[\varphi,T] = 
   \frac{T^4}{2\pi^2} \sum_f  
     N_{c_f} N_{s_f} (-1)^{2 s_f} I_f(a_f), \label{finitetemp}
\label{tmpeff}
\end{eqnarray}
where $I_f(a_f)$ is 
\begin{eqnarray}
   I_B(a_f) &=& \int_0^\infty dx\,\, x^2 \log[1 - e^{-\sqrt{x^2+a_f^2}}],\,\, 
              ({\rm boson}), \\ 
   I_F(a_f) &=& \int_0^\infty dx\,\, x^2 \log[1 + e^{-\sqrt{x^2+a_f^2}}],\,\, 
              ({\rm fermion}), 
\label{tmpeff2}
\end{eqnarray}
for the loop particle $f$ with $a_f = \tilde{m}(\varphi)_f/T$.
For a given mass parameter set, the critical temperature $T_c$ 
and the critical expectation value $\varphi_c$ ($\neq 0$) 
are obtained as solutions of 
\begin{eqnarray}
 V_{eff}[\varphi_c, T_c] &=& 0, \\
 \frac{\partial}{\partial \varphi} 
V_{eff}[\varphi_c, T_c] &=& 0.  
\end{eqnarray}

The expression by using the high temperature expansion~\cite{hte}
(see Appendix D)  
is useful to see the structure in an analytic way. 
When $\sin(\beta-\alpha)=1$, $m_H^{2}, m_{A}^{2}, m_{H^\pm}^{2} \gg M^2$, $m_{S^\pm}^2 \gg
\mu_{S^\pm}^{2}$ and $m_\eta^2 \gg \mu_\eta^2$, 
the effective potential at finite temperatures 
$V_{eff}[\varphi, T] = V_{\rm tree}[\varphi] 
+ \Delta V[\varphi]
+ \Delta V_T[\varphi,T]$ can be approximately expressed as
\begin{eqnarray}
 V_{eff}[\varphi, T]= D (T^2-T_0^2) \varphi^2 
                     - E T \varphi^3 
                     + \frac{\lambda_T}{4} \varphi^4 + ..., 
\end{eqnarray}
where 
\begin{eqnarray}
D &\simeq& \frac{1}{24 v^2} \left(6 m_W^2 + 3 m_Z^2 + 6 m_t^2 
 + \frac{7}{2}m_h^2 + m_H^2+m_{A}^2+2m_{H^\pm}^2
                            %\right.\nonumber\\
% &&\left.
    + 2m_{S^\pm}^2+m_\eta^2 \right),\\
T_0^2 &\simeq& \frac{1}{2D}\left(\frac{m_h^2}{2}-4 B v^2 \right),\\
E &\simeq& \frac{1}{12 \pi v^3} (6 m_W^3 + 3 m_Z^3  
 + 2 m_{H^\pm}^3 + m_H^3 +  m_A^3 + 2
 m_{S^\pm}^3 + m_\eta^3) ,\\
\lambda_T &\simeq& 
 \frac{m_h^2}{2 v^2} 
 \left[ 1  
- \frac{1}{8\pi^2 v^2 m_h^2} 
 \left\{ 
+  6 m_W^4 \log \frac{m_W^2}{\alpha_B^{} T^2}  
+  3 m_Z^4 \log \frac{m_Z^2}{\alpha_B^{} T^2}  
- 12 m_t^4 \log \frac{m_t^2}{\alpha_F^{} T^2}  \right.\right.\nonumber\\
&&+  \frac{9}{4} m_h^4 \log \frac{m_h^2}{\alpha_B^{} T^2}  
+  2 m_{H^\pm}^4 \log \frac{m_{H^\pm}^2}{\alpha_B^{} T^2}  
+  m_H^4 \log \frac{m_H^2}{\alpha_B^{} T^2}  
+  m_A^4 \log \frac{m_A^2}{\alpha_B^{} T^2}  \nonumber\\
&&
\left.\left.
       +  2 m_{S^\pm}^4 \log \frac{m_{S^\pm}^2}{\alpha_B^{} T^2}
%\nonumber\\  
%&& 
+  m_{\eta}^4 \log \frac{m_{\eta}^2}{\alpha_B^{} T^2}  
\right\}   
\right],
\end{eqnarray}
where $\log\alpha_B^{}=2 \log 4 \pi - 2 \gamma_E^{}$ and 
$\log\alpha_B^{}=2 \log \pi - 2 \gamma_E^{}$, and 
\begin{eqnarray}
B&\simeq&\frac{1}{64\pi^2v^4}\left\{
6m_W^4+3m_Z^4-12m_t^4+\frac{9}{4}m_h^4+
m_{H}^4 +m_{A}^4
%\right.\nonumber\\
%&&\left.
+2 m_{H^\pm}^4
+2 m_{S^\pm}^4
+m_{\eta}^4
\right\}.
\end{eqnarray}
We analytically obtain the critical temperature $T_c$ and the order
parameter $\varphi_c^{}$ at $T_c$ as  
\begin{eqnarray}
T_c = T_0 \frac{1}{\sqrt{1 - \frac{E^2}{\lambda_{T_c}^{} D}}}, \hspace{4mm}
\varphi_c = \frac{2 E T_c}{\lambda_{T_c}^{}}.
\end{eqnarray}

In order to satisfy the sphaleron decoupling condition in the broken
phase, it is required that~\cite{sph-cond}  
\begin{eqnarray}
  \frac{\varphi_c}{T_{c}} \gsim 1.  \label{sph}
\end{eqnarray}
In terms of high temperature description of the effective potential,
Eq.~(\ref{sph}) implies the condition  
\begin{eqnarray}
 \frac{2 E}{\lambda_{T_c}} \gsim 1. \label{sph2}
\end{eqnarray}
A large value of $E$ is important for the strongly first-order
EWPT~\cite{ewbg-thdm2}, for which only the bosonic loop effects can
contribute.
In the case of the SM, the value of the coefficient $E$ mainly comes
from loop contributions of weak gauge bosons, so that
the above relation gives the upper bound on $m_h^{}$ which is
much below the lower bound from the LEP experiment.
However, in some new physics models, the situation can be improved due
to the contributions of additional bosonic fields which couple to
$h$. In our model, there are many additional scalars running in the
loop so that a larger $E$ can be easily realized. Consequently, the
first order EWPT is possible without contradiction with the LEP data
in a wide region of the parameter space.

\begin{figure}[t]
\begin{center}
  \epsfig{file=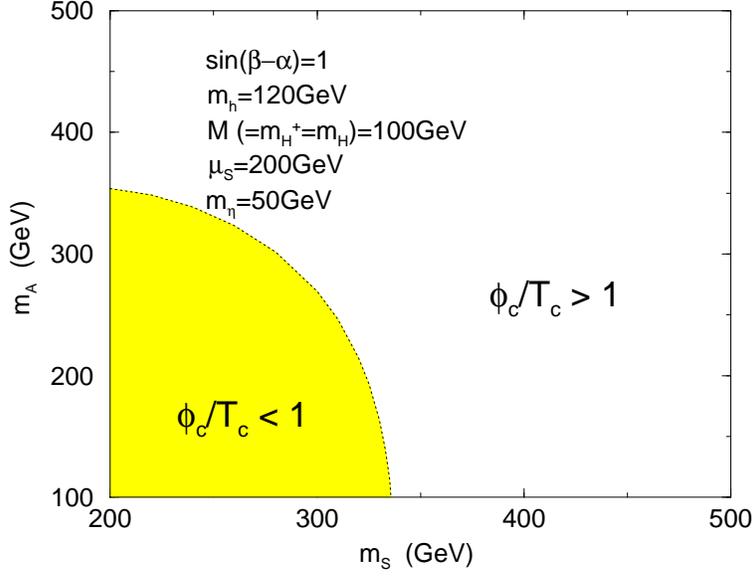,width=10.cm}
\end{center}
\vspace{-2mm}  \caption{The region of strong first order EWPT
 is shown in the $m_A^{}$-$m_{S^\pm}^{}$ plane.
 Shaded (yellow colored) area is excluded by the condition in Eq.~(\ref{sph}).
 The invariant mass parameters $M$ and $\mu_S^{}$ are
 taken as $M = 100$ GeV and $\mu_S^{} = 200$ GeV.}
 \label{1optAS}
\end{figure}
In FIG.~\ref{1optAS}, the allowed region under the condition of
Eq.~(\ref{sph2}) is shown for 
$m_h \simeq 120$ GeV, $m_H^{} \simeq m_{H^\pm}^{} (\simeq M) \simeq 100$
GeV and $\sin(\beta-\alpha)\simeq 1$.
The invariant mass parameters $M$ and $\mu_S^{}$ in Eqs.~(\ref{invmass_ma}) and
(\ref{invmass_ms}) are taken as $M = 100$ GeV  and $\mu_S^{} = 200$ GeV.
The condition is satisfied
when $m_{S^{\pm}}^{} \gsim 360$ GeV for $m_A^{} \simeq 100$ GeV and
when $m_{A}^{} \gsim 340$ GeV for $m_S^{} \simeq 200$ GeV.
If we take larger values for $M$ ($< m_A^{}$) and $\mu_S^{} <
m_{S^\pm}^{}$, then the area of the allowed regions becomes smaller:
in order to obtain similar magnitude of non-decoupling effects a heavier
$A$ or $S^\pm$ is favored.
The result is not sensitive to $\tan\beta$.
We note that the approximation with high temperatures becomes worse for
larger $M$ ($M \gg T$) but the conclusion here is not qualitatively changed
for our parameter set even when we evaluate it without high temperature
expansion.

\section{Phenomenology}

\subsection{Constraints on the parameters from the current data}

Before going to the discussion on the prediction of the model,
we here summarize the constraints on the model from the following current experimental
data and the theoretical requirements.\\

\noindent 
\underline{The $\mu \to e \gamma$ results:}

The constraint from the data for $\mu \to e \gamma$ and the natural
generation of the tiny neutrino masses require
that $m_{N_R^\alpha}^{} > {\cal O}(1)$ TeV and $m_{S^\pm}$ is several hundred GeV. \\

\noindent 
\underline{Tiny neutrino masses with ${\cal O}(1)$ coupling constants:}

We require no unnatural fine tuning on the coupling constants $\kappa$
and $h_e^\alpha$; i.e. $\kappa \sim h_e^\alpha \sim {\cal O}(1)$.
 In this case, the data prefer
$m_{H^\pm}^{} \lsim 100$ GeV,  $m_{S^\pm}^{} ={\cal O}(100)$ GeV and
$\kappa\tan\beta \simeq 30$.
We do not explain the mass hierarchy among charged leptons, so that
we consider scales of $h_i^\alpha$ to satisfy
${\cal O}(h_i^\alpha y_\ell^i) \sim 10^{-5} \sim {\cal O}(y_e)$.  \\

\noindent 
\underline{The data from the LEP and from the Tevatron:}

Throughout this paper, we take 
\begin{eqnarray}
   m_{H^\pm}^{} = m_H^{}, \hspace{2mm}{\rm and} \hspace{2mm}
    \sin(\beta-\alpha)=1. 
 \end{eqnarray}
In this case, the Higgs potential has
the global custodial $SU(2)_V$ symmetry. 
The bound from the LEP precision measurement especially those
for electroweak rho parameter is easily satisfied.
 The scalar $h$ plays the same role as the SM Higgs boson.
 The other scalar bosons do not contribute to the rho parameter. 
 Therefore, we have similar mass bound on $h$ to that on the SM Higgs boson from the LEP
 data; i.e., $m_h=114$-$160$ GeV.  
 The LEP lower bound for charged Higgs boson mass should be
 $m_{H^\pm} \gsim 100$ GeV, which must also be respected.
 Combining this with the requirement of natural generation of
 tiny neutrino mass, we obtain
 $m_{H^\pm}^{} = m_H^{}\sim 100$ GeV. A little bit larger values are
 also allowed when we take small fine tuning for the coupling constants.\\

\noindent 
\underline{The $b\to s \gamma$ results:}

As discussed in previous section, the results from $b\to s\gamma$
strongly constrain the mass of charged Higgs bosons in the Type-II Yukawa interaction in THDMs. It gives the
lower bound as $m_{H^\pm}^{} \gsim 295$ GeV~\cite{Ref:bsgNNLO}.
In order to avoid this constraint and in order to have a compatible
value of $m_{H^\pm}^{}$ with the neutrino data, we choose Type-X Yukawa
interaction in our model.  \\
 
\noindent 
\underline{The $g-2$ results:}

The predicted muon $g-2$ in this model with the above parameter sets of new particle mass and 
Yukawa couplings to reproduce neutrino mass is far below the current experimental bound~\cite{mg-2}.
In addition, the electron $g-2$ is also below the current bounds
even for order unity couplings $h_e$ because it is suppressed by the
electron mass~\cite{eg-2}. 
Thus, there are no significant constraints on the model from $g-2$. \\

\noindent 
\underline{The dark matter data:}

The DM thermal relic abundance $\Omega_{\eta}h^2 \simeq 0.1$ (WMAP)
requires $m_\eta \lesssim 65$ GeV for $m_S \gtrsim 400$ GeV, 
or $m_{\eta}\simeq m_h/2 $ or $m_H/2$. 
The null result of direct DM searches corresponds 
to couplings $\sigma \lesssim 10^{-2}$ for $m_{\eta} \gtrsim 20$ GeV.\\

\noindent 
\underline{Strong first order phase transition:}

We need non-decoupling property in the Higgs sector to realize
the strong first order phase transition of electroweak symmetry.
As, the mass of $H$ and $H^\pm$ are constrained to be around 100 GeV,
these fields cannot give such strong non-decoupling loop effects.
We consider the case where $A$ and $S^\pm$ are relatively heavy
so that the sufficient non-decoupling loop effects can be generated.
Concretely, we may consider $m_A=100-200$ GeV, and $m_{S^\pm}^{} \gsim
400$ GeV.

\subsection{Predictions}

We now show phenomenological predictions in the scenario in~(\ref{scenario}) in
order.  Here we mainly work on the  scenario which can simultaneously solve the above three issues 
under the data~\cite{lep-data,MEGA,bsgamma}:
\begin{eqnarray}
 \begin{array}{ll}
 \sin(\beta-\alpha) =  1, &\!\! \kappa \tan\beta \simeq 30, \\
 m_h = 120 {\rm GeV},     &\!\! m_H^{} \simeq m_{H^\pm} (\simeq M) \simeq 100 {\rm GeV},    \\
 m_A \sim 100-200 {\rm GeV},     &\!\! m_{S^\pm}^{} \sim 400{\rm GeV},\\
 m_{\eta} = 40-65 {\rm GeV}, &\!\! m_{N_R^{1}} = m_{N_R^{2}} =  3 {\rm TeV}.\\
  \end{array} \label{scenario}
\end{eqnarray}
By this scenario, the model can explain the origin of tiny neutrino mass and
mixing, the origin of DM, and the strongly first-order phase transition which is
necessary for successful electroweak baryogenesis.
This can be realized without assuming unnatural hierarchy among the
coupling constants. All the masses are between
${\cal O}(100)$ GeV and ${\cal O}(1)$ TeV. 
They are indicated by the data, so that
the model has a predictive power.\\

\noindent 
\underline{Invisible decays of the Higgs boson to a DM pair:}

\begin{figure}[t]
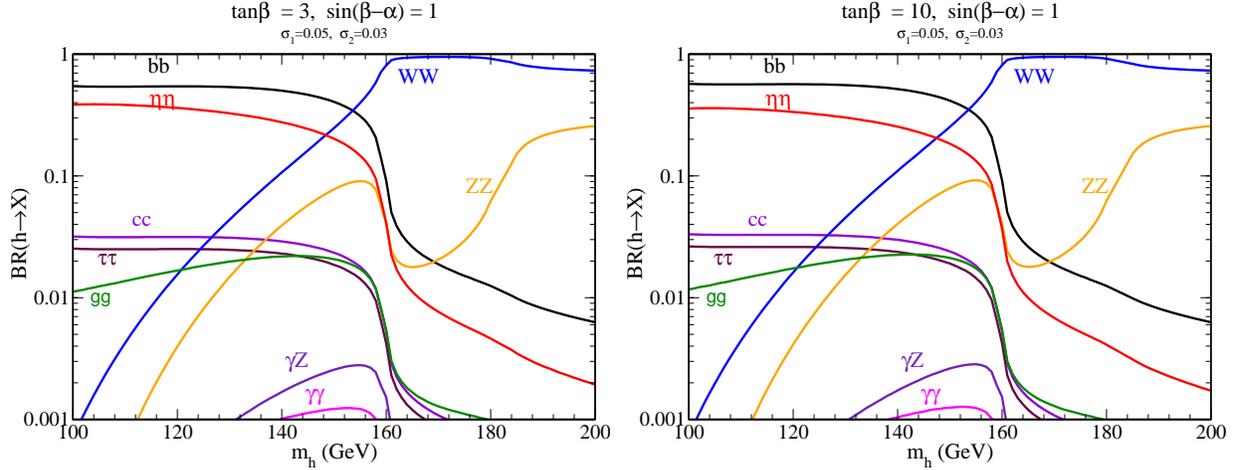

\begin{minipage}{0.49\hsize}
\includegraphics[width=8cm]{AKS_hdecay3.eps}
\end{minipage}
\begin{minipage}{0.49\hsize}
\includegraphics[width=8cm]{AKS_hdecay10.eps}
\end{minipage}
  \caption{Branching ratios of the SM like Higgs boson $h$ for
 $\tan\beta=3$ (left) and $\tan\beta=10$ (right) for
 $\sin(\beta-\alpha)=1$ and $m_\eta = 48$ GeV.
 The coupling constants are taken to be $\sigma_1=0.05$, 
 $\sigma_2=0.03$, $\rho_1=1.0$ and $\rho_2=1.0$.
 }
  \label{fig:br_hetaeta}
\end{figure}
The scalar boson $h$ is the SM-like Higgs boson, whose coupling constants
to the SM particles coincide with that of the SM Higgs boson at the tree
level. Therefore, its production mechanisms are the same as those in the SM.
However, when $m_\eta < m_h/2$ it decays into a DM  pair $\eta\eta$, 
namely the invisible decay of Higgs as a common prediction of 
singlet scalar DM (such as  Higgs portal model)~\cite{InvisibleDecayByDM}, whose decay rate is given by
 \begin{eqnarray}
      \Gamma(h\to \eta\eta)\! = \!\frac{v^2}{32\pi
      m_h^{}}\sqrt{\!1\!-\!\frac{4m_\eta^2}{m_h^2}\!}
      |\sigma_1\cos^2\beta\!+\!\sigma_2\sin^2\beta|^2.
 \end{eqnarray}
The branching ratio in our model is shown in
Fig.~\ref{fig:br_hetaeta}.
When $m_h=120$ GeV and $m_\eta=48$ GeV, the branching ratio of $h\to \eta\eta$ amounts to
about 36 \% (34\%) for $\tan\beta=3$ ($10$).
For $m_h=120$ GeV and $m_\eta=55$ GeV, it is about 25 \% (22 \%) for
$\tan\beta=3$ ($10$).
For heavier $h$ than 120 GeV, the branching ratio of $h\to\eta\eta$ is
smaller because the mode $h\to WW^\ast$ becomes larger. 
The decay rate is strongly related to the DM abundance: see Eq.~(\ref{annihiration}) 
we may be able to reconstruct at least the annihilation into $f\bar{f}$,
so that our DM scenario can be testable at the Large Hadron Collider
(LHC).\\

\noindent 
\underline{DM searches:}

        \begin{figure}[t]
        \begin{center} 
           \epsfig{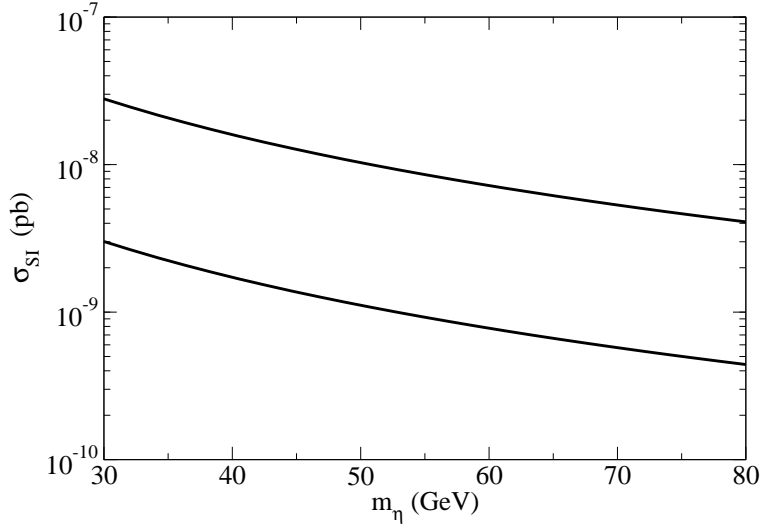}
        \end{center}
        \vspace{-2mm}
        \caption{The cross section for the scattering process with a
         proton at the direct search experiment. The upper (lower) curve
         corresponds to the parameter $\sigma_2=0.03$ ($0.01$). The
         other parameters are taken to be $m_h=120$ GeV, $m_H=100$ GeV, $\sigma_1=0.05$
         and  $\tan\beta=10$.  }  \label{fig:standardSI}
        \end{figure}
 Our DM candidate
$\eta$  is potentially detectable by direct and indirect DM searches.
        For the coupling constants $\sigma_{i} \sim 10^{-2}$ and the
        mass $m_{\eta} \sim 50$ GeV, 
        the typical SI cross section is of order of $10^{-8}$ pb 
        and within the reach of near future direct search experiments such as
        superCDMS and XMASS~\cite{xmass}.
        In Fig.~\ref{fig:standardSI}, the SI cross section of DM 
        for a proton $\sigma_p^{SI}$ is plotted.
        In addition, indirect searches by observing cosmic rays which would be 
        a signal of DM annihilation is also promising. 
        It is regarded that a line gamma is the smoking gun of DM annihilation.
        $\eta$ has a large cross section of annihilation into two photon.
        Although this is a one-loop process, 
        the cross section is not so much suppressed because of 
        a large coupling $\kappa$ and light particles 
        inside loops ($S^\pm$ and $H^\pm$).
        As a similar case, 
        one may recall the inert Higgs doublet model where the $W^+$-boson 
        loop gives a sizable contribution~\cite{Gustafsson:2007pc}.
        The FERMI satellite~\cite{Gehrels:1999ri}, used to be called GLAST,
        sensitivity with the solid angle of detector $\Delta\Omega\ \sim 10^{-5}$
        is expected to be $\gtrsim 10^{-9} {\rm GeV cm^{-2} s^{-1}}$ for ${\cal O}(10)$ GeV energy.
        Line gamma-ray emissions could be observed 
        for the Navarro, Frenk, and White (NFW) density profile~\cite{Navarro:1996gj} with
        the astrophysics dependent dimensionless function $J \sim 10^4$,
        because the flux originated from $\eta$ annihilation is expected to be as large as
\begin{eqnarray}
E^2 \Phi = 5 \times 10^{-8} \left(\frac{50 {\rm GeV}}{m_{\eta}}\right)\left(\frac{\sigma v}{10^{-9} {\rm GeV}^{-2}}\right)
  J \Delta\Omega\, {\rm [GeV cm^{-2} s^{-1}]}
  \label{LineGammaFlux}
\end{eqnarray}
        with $E = m_{\eta}$.\\

\noindent 
\underline{The non-decoupling effect of $S^\pm$ on the $hhh$ coupling:}

\begin{figure}[t]
\begin{center}
  \epsfig{file=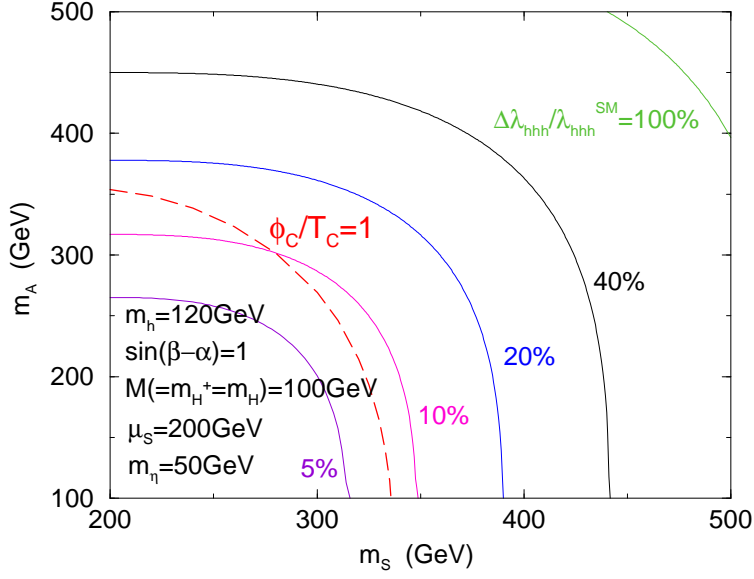,width=10.cm}
\end{center}
\vspace{-2mm}  \caption{
 The contour plots represent deviations in the $hhh$ coupling
 from the SM value. For the invariant mass parameters, $M=100$ GeV and
 $\mu_S^{} = 200$ GeV are taken. The region of the strongly first order phase
 transition at this parameter choice is also shown.}
 \label{hhh}
\end{figure}

        As we have seen the condition for the first order
        electroweak phase
        transition requires non-decoupling property for $S^\pm$ or $A$.
         It is known that such non-decoupling property affects physics
         of the renormalized $hhh$ coupling.
        In terms of the renormalized mass parameter $m_h$, the one-loop 
corrected tri-linear coupling is obtained from the effective potential~\cite{nondec}, 
\begin{eqnarray}
 \lambda_{hhh}^{eff} \!\!&=&\!\! \frac{3 m_h^2}{v}
      \left\{ 1  
              + \frac{m_{H}^4}{12 \pi^2 m_h^2 v^2} 
                         \left(1 - \frac{M^2}{m_H^2}\right)^3 
              + \frac{m_{A}^4}{12 \pi^2 m_h^2 v^2} 
                         \left(1 - \frac{M^2}{m_A^2}\right)^3
      \right. \nonumber \\
 &&           + \frac{m_{H^\pm}^4}{6 \pi^2 m_h^2 v^2} 
                         \left(1 - \frac{M^2}{m_{H^\pm}^2} \right)^3 
              + \frac{m_{S^\pm}^4}{6 \pi^2 m_h^2 v^2} 
                         \left(1 - \frac{\mu_S^2}{m_{S^\pm}^2}\right)^3
                         \nonumber \\
&&\left.
\!\!\!\!\!\!\!\!\!\!\!\!
              + \frac{m_{\eta}^4}{12 \pi^2 m_h^2 v^2} 
                         \left(1 - \frac{\mu_\eta^2}{m_{\eta}^2}\right)^3
              - \frac{N_{c_t} m_t^4}{3 \pi^2 m_h^2 v^2} + 
              {\cal O} \left(\frac{p^2_i m_\Phi^2}{m_h^2 v^2},
                           \;\frac{m_\Phi^2}{v^2},
                           \;\frac{p^2_i m_t^2}{m_h^2 v^2},  
                           \;\frac{m_t^2}{v^2}  \right)
      \right\}. \label{m4THDM}    
\end{eqnarray}
The deviation from the SM prediction 
\begin{eqnarray}
 \lambda_{hhh}^{eff}({\rm SM}) \!\!=\!\! \frac{3 m_h^2}{v}
      \left\{ 1  
              - \frac{N_{c_t} m_t^4}{3 \pi^2 m_h^2 v^2} + 
              {\cal O} \left(\frac{p^2_i m_t^2}{m_h^2 v^2},  
                           \;\frac{m_t^2}{v^2}  \right)
      \right\}, \label{m4SM}    
\end{eqnarray}
is defined by $\Delta \lambda_{hhh}/\lambda_{hhh}^{eff}({\rm SM})$, where 
$\Delta \lambda_{hhh}
=\lambda_{hhh}^{eff}
-\lambda_{hhh}^{eff}({\rm SM})$.
        The quantum effect on the $hhh$ coupling amounts to more
        than 10-20 \% (see FIG.~\ref{hhh}), due to the non-decoupling
        property of $A$ and
        $S^\pm$, which is required for successful baryogenesis~\cite{ewbg-thdm2,ewbg-thdm3}.
        Thus, it should be testable  at the International Linear
        Collider (ILC)~\cite{hhh-measurement,gamgamhh2,gamgamhh}.

  We note that in addition to the non-decoupling effect on the $hhh$
  coupling the non-decoupling property of the charged singlet scalars
  $S^\pm$ can be further tested by measuring
  $\gamma\gamma h$ vertices~\cite{zee-ph}. \\

 \noindent 
\underline{Phenomenology of the Type-X Yukawa interaction:}
       
Because of the Type-X Yukawa coupling defined in
        Eq.~(\ref{eq:yukawa1})~\cite{typeX,typeX2}, the 
        phenomenology for the extra Higgs bosons is 
        different from that in Type I or Type II especially for $\tan\beta\gsim 2$.
         In our model $H$ and $A$ ($H^\pm$) can predominantly decay into
        $\tau^+\tau^-$ ($\tau^\pm\nu$), while  in the Type II THDM 
        the main decay modes of $H$ and $A$ ($H^\pm$) 
        are $b\bar b$ ($\tau \nu$) when the masses are around 100 GeV
        and $\sin(\beta-\alpha)=1$.

        Recently, the phenomenology in
        Type-X THDM has been studied in the similar parameter choice in Ref.~\cite{typeX}. 
        The physics of Type-X Yukawa interaction can be tested at the
        LHC with the low luminosity (30 fb$^{-1}$) 
        via the single direct (associated) production processes $gg \to
        \Phi \to \tau^+\tau^-$ and $\mu^+\mu^-$ ($pp\to b\bar b\Phi \to
        b\bar b \tau^+\tau^-$ and $b\bar b \mu^+\mu^-$), where $\Phi=A$
        and $H$ .  
        At high luminosity (300 fb$^{-1}$), $pp\to W^\ast \to A H^\pm$ and $pp\to W^\ast
        \to H H^\pm$~\cite{wah} can also be used to discriminate between
        the types of Yukawa interaction when the masses are
        not too large.
        In the minimal supersymmetric standard model (Type II Yukawa interaction) the main
        signal would be the $b\bar b \tau \nu$ final state while that
        in the Type-X  the final main states would be
        $\tau^+\tau^-\tau^\pm\nu$ or $\mu^+\mu^-\tau^\pm\nu$.   \\

        \noindent 
\underline{The phenomenology of the charged singlet scalar field $S^\pm$:}

\begin{figure}[t]
\begin{center}
  \epsfig{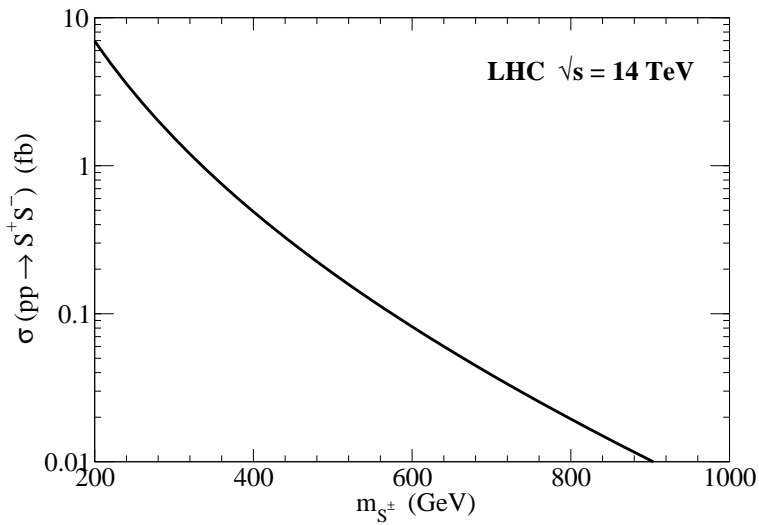}
\end{center}
  \caption{Production cross section of a $S^+S^-$ pair via
 the Drell-Yan process at the LHC ($\sqrt{s}=14$ TeV) as a function of the mass of $S^\pm$.
} 
  \label{fig:SS_LHC}
\end{figure}

The physics of $Z_2$-odd charged singlet 
        $S^\pm$ is important to distinguish this model from the
        other models.
        At the LHC, they are produced in pair via the Drell-Yan
        process~\cite{zee-ph}.
        In Fig.~\ref{fig:SS_LHC}, the cross section for $pp\to
        S^+S^-$ at the LHC is shown as a function of the mass
        $m_{S^\pm}^{}$. 
        The cross section amounts to 0.5 fb for $m_{S^\pm}^{}$, so that
        more than a hundred of the $S^+S^-$ events are produced for the
        integrated luminosity $300$ fb$^{-1}$.
        The produced $S^\pm$ bosons decay as $S^\pm \to H^\pm \eta$, 
        and $H^\pm$ mainly decay into $\tau^\pm \nu$
        due to the Type-X Yukawa coupling when $\tan\beta\gsim 2$~\cite{typeX}. 
        The signal would be a high-energy hadron pair~\cite{hagiwara} with a large missing
        transverse momentum.

\begin{figure}[t]
\begin{minipage}{0.4\hsize}
\includegraphics[width=9cm]{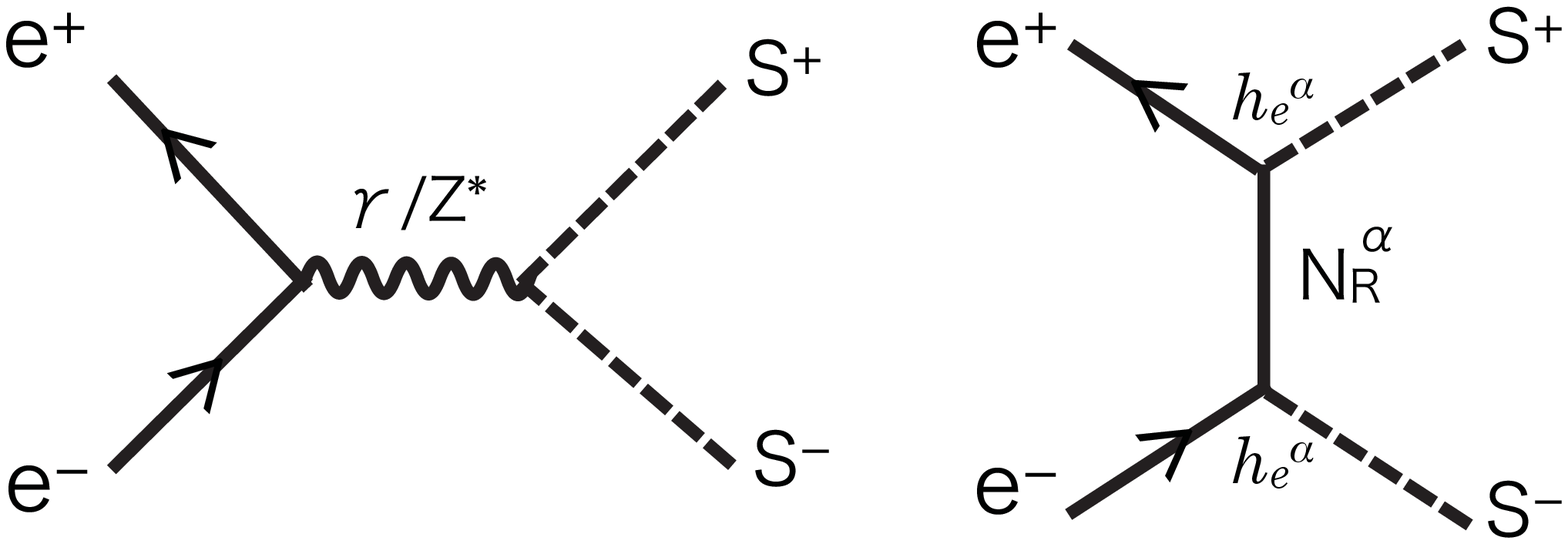}
\end{minipage}
 \hspace{3cm}
\begin{minipage}{0.35\hsize}
\includegraphics[width=4.5cm]{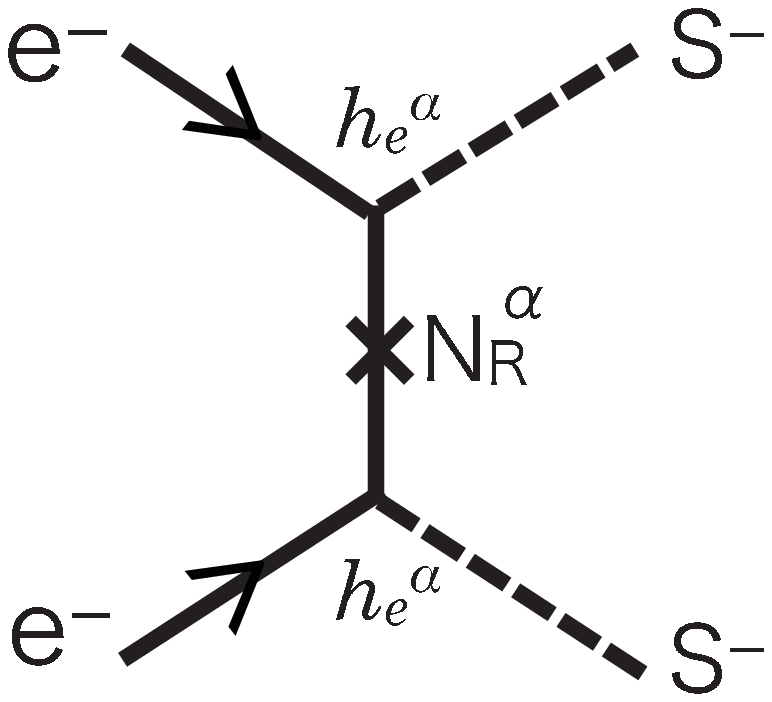}
\end{minipage}\\
\vspace{-2mm}  \caption{Feynman diagrams for the processes of  $e^+e^-\to
 S^+S^-$ (Left) and $e^-e^-\to S^-S^-$ (Right).}
 \label{fig:smsp_diag}
\end{figure}

\begin{figure}[t]
\begin{minipage}{0.49\hsize}
\includegraphics[width=8cm]{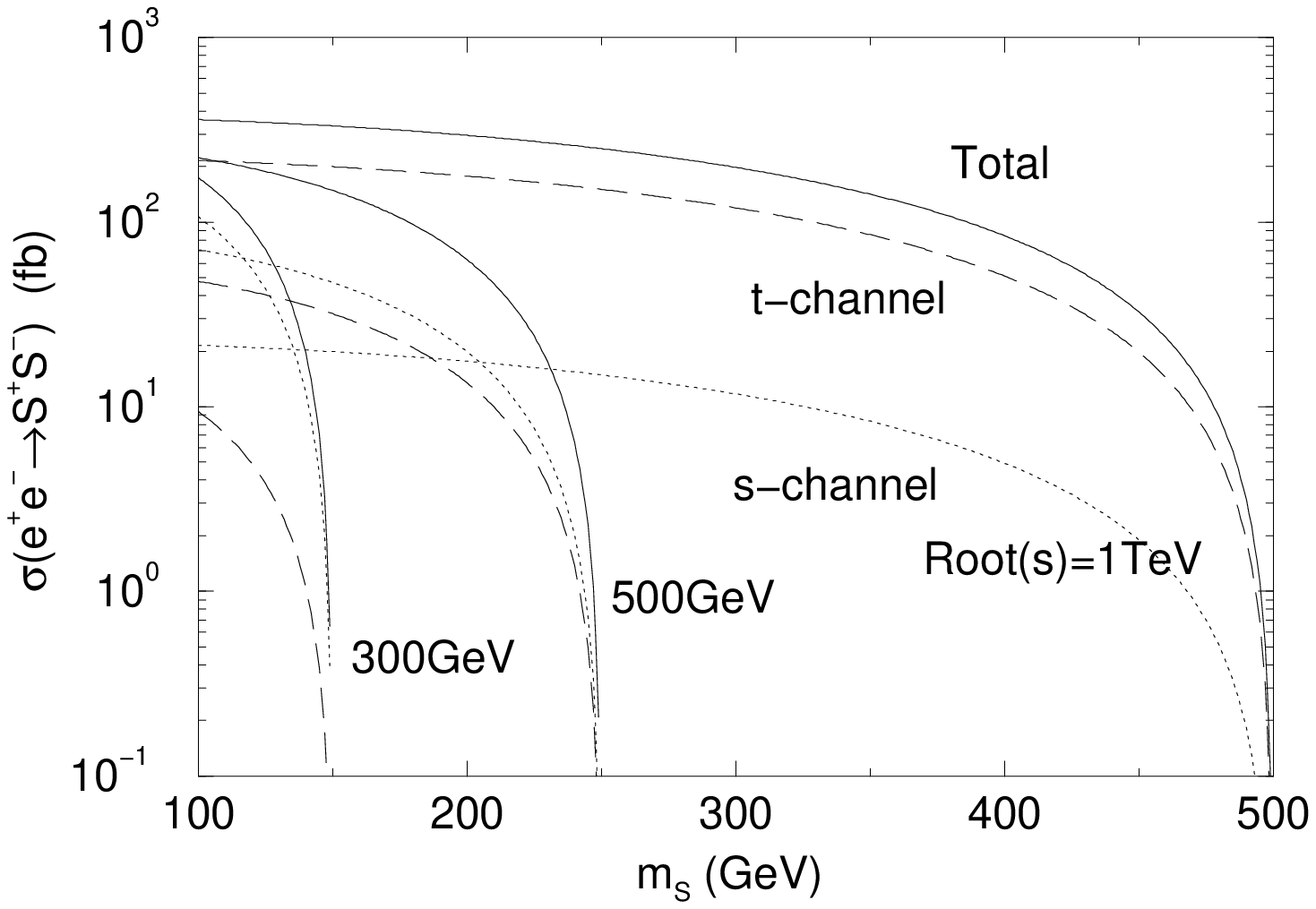}
\end{minipage}
\begin{minipage}{0.49\hsize}
\includegraphics[width=8cm]{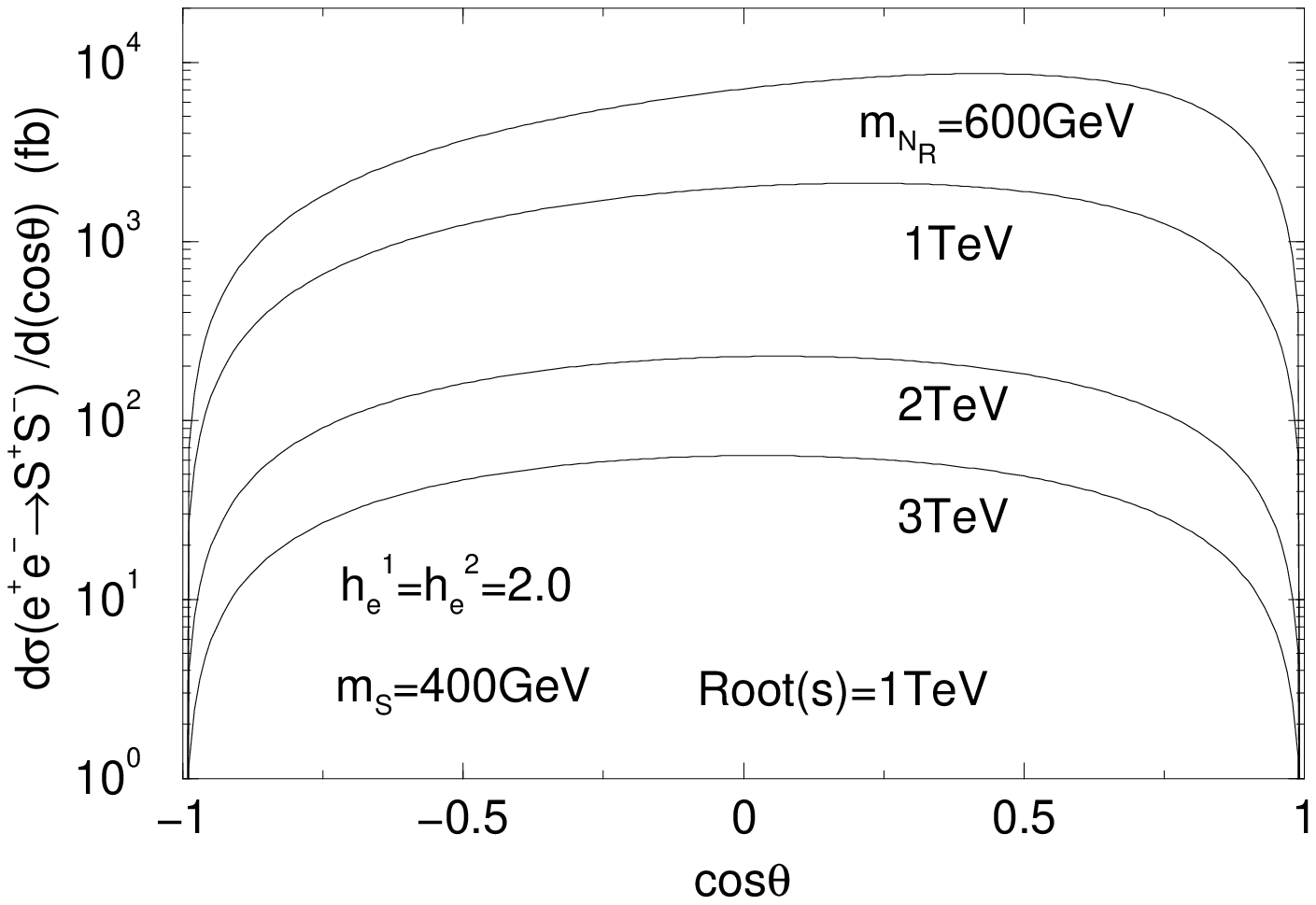}
\end{minipage}\\
  \caption{
 [Left]
 Production cross sections for $e^+e^- \to S^+S^-$ via the s-channel gauge boson
 ($\gamma$ and $Z$) mediation (dotted curve), the t-channel RH-neutrino
 ($N_R^\alpha$) mediation (solid curve), and both contributions (dashed
 curve) for $\sqrt{s}=300$, 500 and 1000 GeV.  The masses for RH neutrinos
 are taken to be $m_{N_R^1}^{}=m_{N_R^2}^{}=3$ TeV.
  [Right]
The angular distribution of the $e^+e^- \to S^+S^-$ at $\sqrt{s}=1$ TeV
    for $m_{N_R^1}^{}(=m_{N_R^2}^{})=0.6$, $1$, $2$ and $3$ TeV. 
In both figures, $h_e^1=h_e^2=2.0$ are taken.} 
  \label{fig:smsp}
   \end{figure}

   The charged singlet scalar bosons $S^\pm$ in our model
   can also be better studied at the ILC via $e^+e^-\to S^+S^-$
   shown in Fig.~\ref{fig:smsp_diag}(Left). 
   The total cross sections
    are shown as a function of $m_{S^\pm}^{}$ for
   several values of the center-of-mass energy $\sqrt{s}$
   in Fig.~\ref{fig:smsp}(Left).
   The other relevant parameters are taken as 
   $m_{N_R^1}^{}=m_{N_R^2}^{}=3$ TeV and $h_e^1=h_e^2=2.0$.
   Both the contributions from the s-channel gauge boson ($\gamma$ and $Z$) mediation and the
   t-channel RH neutrino mediation are included in the calculation.
   The total cross section can amount to about 100 fb for
   $m_{S^\pm}^{}=400$ GeV at $\sqrt{s}=1$ TeV due to the contributions
   of the t-channel RH neutrino-mediation diagrams with ${\cal O}(1)$
   coupling constants $h_e^\alpha$.
   The signal would be a number of energetic tau lepton pairs with
   large missing energies. 
   Although several processes such as $e^+e^-\to W^+W^-$ and $e^+e^-\to H^+H^-$   
   can give backgrounds for this final state,
   we expect that the signal events can
   be separated by kinematic cuts.
In Fig.~\ref{fig:smsp}(Right), the dependences on the scattering angle
in the differential cross section $d\sigma/d(\cos\theta)$
are shown. For $m_{N_R^1}^{}=m_{N_R^2}^{}=3$ TeV, the
special behavior in the angular distribusion is more insensitive than the
cases with lighter values for $m_{N_R^{\alpha}}^{}$. 

\begin{figure}[t]
 \begin{center}
  \epsfig{file=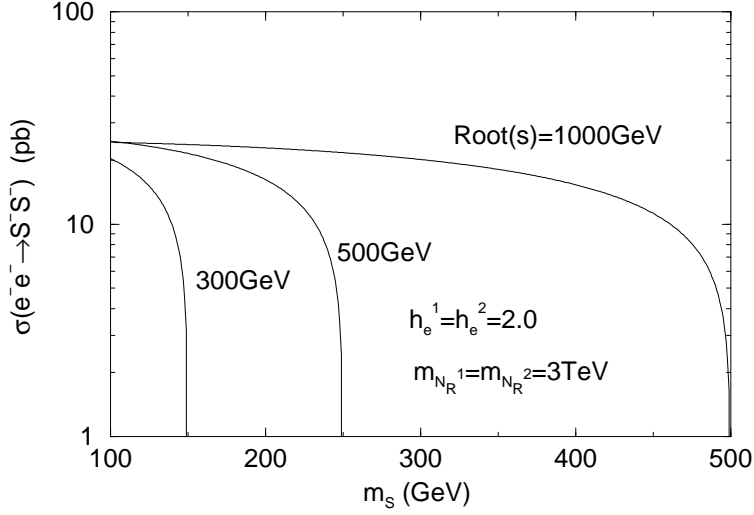,width=10.cm}
\end{center}
  \caption{ Production cross sections for $e^-e^- \to S^-S^-$ via the t-channel RH-neutrino
 ($N_R^\alpha$) mediation (solid curve) for $\sqrt{s}=300$, 500 and 1000 GeV.  
 The masses for RH neutrinos and the relevant coupling constants 
 are taken to be $m_{N_R^1}^{}=m_{N_R^2}^{}=3$
 TeV and $h_e^1=h_e^2=2.0$, respectively.} 
  \label{fig:smsm}
\end{figure}

   Finally, there is a further advantage in testing our model at the
   $e^-e^-$ collision option of the ILC, where 
   the dimension five operator $e^- e^- S^+ S^+$, which appears in
   the sub-diagram of the three loop induced masses
   of neutrinos in our model, can be directly measured.
   The production cross section for $e^- e^- \to S^-S^-$ [t-channel
   $N_R^\alpha$ mediation: see Fig.~\ref{fig:smsp_diag}(Right)] is given by
  \begin{eqnarray}
    \sigma(e^- e^- \to S^-S^-) = \int_{t_{\rm min}}^{t_{\rm max}}
     dt 
\frac{1}{128\pi s}  
      \left|\sum_{\alpha=1}^2
       (|h_e^\alpha|^2 m_{N_R^\alpha})
       \left(\frac{1}{t-m_{N_R^\alpha}^2}+\frac{1}{u-m_{N_R^\alpha}^2}\right)\right|^2. 
   \end{eqnarray}
  Because of the structure of our model that the tiny neutrino masses are
  generated at the three loop level, the magnitudes of  $h_e^\alpha$
  ($\alpha=1,2$)
  are of ${\cal O}(1)$, by which the cross section becomes very large.
  Furthermore, thanks to the Majorana nature of the t-channel diagram, 
  we obtain a much larger cross section in the $e^-e^-$ collision than at
  the $e^+e^-$ collision when $m_{N_R^\alpha}^{2} \gg s$.
  The cross section can be as large as 10 pb for $m_{S^\pm}^{}=400$ GeV for
  $\sqrt{s}_{e^-e^-} = 1$ TeV, $m_{N_R^{1}}^{}=m_{N_R^{2}}^{}=3$ TeV and
  $h_e^1=h_e^2=2.0$: see Fig.~\ref{fig:smsm}. 
  The backgrounds are expected to be much less than the $e^+e^-$ collision. 

      We emphasize that a combined study for these processes
      would be an important test  for our model, in
  which neutrino masses are generated at the three loop level by the 
  $Z_2$ symmetry and the TeV-scale RH neutrinos\footnote{Unlike our
      model, in the model in
      Ref.~\cite{knt}, the coupling constants corresponding to our
      $h_e^\alpha$ are small and instead those to $h_\mu^\alpha$ are
      ${\cal O}(1)$, so that its Majorana structure is not easy to test
      at $e^-e^-$ collisions.}. 
  In the other radiative seesaw models in which the neutrino masses are
  induced at the one-loop level with RH neutrinos, the corresponding
  coupling constants to our $h_e^\alpha$ couplings are necessarily
  one or two orders of magnitude smaller to satisfy the neutrino data, so that the cross
  section of the
  t-channel RH neutrino mediation processes are small due to the suppression 
  factor $(h_e^\alpha)^4$.\\

        \noindent 
\underline{Lepton flavor violation:}

Finally, the couplings $h_i^\alpha$ cause lepton flavor violation 
     such as $\mu\to e\gamma$, depending on $m_{N_R^{}}$. If 
     such a phenomenon is observed at future experiments~\cite{meg},
     we could obtain information on $m_{N_R}^{\alpha}$. 
     In addition, our model predicts $B(\mu\to e\gamma)\gg B(\tau\to
     e\gamma) \gg B(\tau\to \mu\gamma)$. \\
     
In summary, the possible scenario in this model provides discriminative 
     phenomenological characteristics so that it can be tested at future experiments.\\

     We have discussed the various features of this model neglecting the CP
violating phases in the Higgs sector, which are crucial for generating
baryon number at the EWPT.
We comment on the case with the CP violating phases.
Our model includes the THDM, so that
the same generation mechanism can be applied in evaluation of produced
baryon number asymmetry at the
EWPT unless $\tan\beta$ is too large~\cite{ewbg-thdm,ewbg-thdm3}.
For a larger value of $\tan\beta$, the constraint from the
EDM data would be more serious. 
The mass spectrum in the Higgs sector would be changed by including
the CP violating phases, but most of the phenomenological features
discussed above should be conserved with a little modification. 

\section{Discussions and Conclusions}

 We have discussed  the model 
 in which  neutrino oscillation, DM, and baryon asymmetry of 
 the Universe can be simultaneously explained by the TeV-scale physics
 without introducing large fine tuning.
 Tiny neutrino masses are generated at the three loop level due to
 the exact $Z_2$ symmetry, by which stability of the DM
 candidate is also guaranteed. The extra Higgs doublet is
 required not only for the tiny neutrino masses but also for
 successful electroweak baryogenesis.
 The phenomenology of the model has been discussed, 
and it has been found that there are several successful scenarios
under the constraints from the current experimental data.
 The predictions have been discussed in these scenarios
 at the present and future experiments.
 It turns out that the model provides discriminative predictions
 especially in Higgs physics and DM physics, 
 so that it is thoroughly testable in future experiments.
\\

\noindent
{\bf Acknowledgments}

We would like to thank Andrew Akeroyd, Shigeki Matsumoto,  
Yasuhiro Okada, Eibun Senaha and Koji Tsumura for useful discussions.
O.~S. would like to thank the warm hospitality of the theoretical
physics group at the University of Toyama.
This work was supported, in part, by 
Grant-in-Aid of the Ministry of Education, 
Culture, Sports, Science and Technology, Government of
Japan, No.~18034004 (SK), and 
DOE Grant No. DE-FG02-94ER-40823 (OS). 

\newpage
%\appndix

\section*{APPENDICES}

\subsection{SM-like limit}

Especially, for the case of the SM like limit ($\sin(\alpha-\beta)=-1$~\cite{Ref:SMlike}, namely
$\alpha=\beta-\pi/2$, ($\sin\alpha=-\cos\beta$ and
$\cos\alpha=\sin\beta$)),  $h$  becomes the SM-like Higgs boson and 
$H$ decouples from the gauge fields (with respect to the three point couplings).
In this case, the Yukawa interactions are 
\begin{eqnarray}
 {\cal L}_Y^{\rm Quarks} &=& - m_t \bar t t + \frac{m_t\cot\beta}{v} H \bar t
  t - \frac{m_t}{v} h \bar t t  \nonumber\\
         && - m_b \bar b b + \frac{m_b\cot\beta}{v} H \bar b b
  - \frac{m_b}{v} h \bar b b  \nonumber\\
 && + \frac{m_t}{v} z \bar t i \gamma_5 t + \frac{m_t}{v}\cot\beta A
  \bar t i \gamma_5 t  \nonumber\\
 && -\frac{m_b}{v} z \bar b i \gamma_5 b - \frac{m_b}{v} \cot\beta A \bar
  b i \gamma_5 b  \nonumber\\
 && + \omega^- \bar b \left[ \frac{\sqrt{2}m_t}{v}
                       \frac{1+\gamma_5}{2}-\frac{\sqrt{2}m_b}{v}
                       \frac{1-\gamma_5}{2}\right] t + {\rm h.c.}  \nonumber\\
 && + H^- \bar b \left[ \frac{\sqrt{2}m_t\cot\beta}{v} 
                       \frac{1+\gamma_5}{2}-\frac{\sqrt{2}m_b\cot\beta}{v}
                       \frac{1-\gamma_5}{2}\right] t + {\rm h.c.}  
\end{eqnarray}
and 
 \begin{eqnarray}
{\cal L}_Y^{\rm Leptons}&=&
    +\frac{\sqrt{2}m_\tau}{v}\bar \nu \frac{1+\gamma_5}{2} \tau \omega^+ +
    {\rm h.c.} \nonumber\\
 &&   -\frac{\sqrt{2}m_\tau \tan\beta}{v} \bar \nu\frac{1+\gamma_5}{2} \tau
    H^++{\rm h.c.} \nonumber\\
 && -m_\tau \bar \tau \tau -\frac{m_\tau\tan\beta}{v}  H
    \bar \tau \tau - \frac{m_\tau}{v}  h
    \bar \tau \tau \nonumber\\
    && -\frac{m_\tau}{v} z \bar \tau i \gamma_5 \tau +
      \frac{m_\tau}{v}\tan\beta A \bar \tau i\gamma_5 \tau.
 \end{eqnarray}

In the decoupling limit ($\sin(\alpha-\beta)=-1$), 
the masses of scalar fields are expressed as
\begin{eqnarray}
  m_h^2 &=& M_{11}^2 = (\lambda_1\cos^4\beta+\lambda_2\sin^4\beta+2\lambda\cos^2\beta\sin^2\beta)v^2, \\
  m_H^2 &=& M_{22}^2 = M^2 + (-\lambda_1\cos^2\beta+\lambda_2 \sin^2\beta+\lambda
   \cos 2\beta)\cos\beta\sin\beta v^2, \\
  m_{H^\pm}^2 &=& M^2-\frac{\lambda_4+\lambda_5}{2} v^2, \\ 
  m_A^2 &=& M^2 -\lambda_5 v^2, \label{invmass_ma} \\
  m_{S^\pm}^2 &=& \mu_S^2 + (\rho_1 \cos^2\beta + \rho_2 \sin^2\beta)\frac{v^2}{2},\label{invmass_ms} \\
  m_{\eta}^2 &=& \mu_\eta^2 + (\sigma_1 \cos^2\beta + \sigma_2 \sin^2\beta)\frac{v^2}{2}.
\end{eqnarray}

\subsection{Neutrino mass matrix}\label{app:calcF}

In order to compute the three loop induced neutrino mass matrix
$M_{ij}^\nu$ in Eq.~(\ref{eq:mij}), we start from
the calculation of the effective vertex of
$\ell_R(p_1)$-$\ell_R(p_2)$-$H^+(p_3)$-$H^+(p_4)$ comes from 
the loop by $Z_2$-odd particles, 
$i T(p_1,p_2,p_3,p_4)$, where $p_1$, $p_2$, $p_3$ and $p_4$ are
 incoming momenta of $\ell_R^i(p_1)$, $\ell_R^j(p_2)$, $H^+(p_3)$,
 and $H^+(p_4)$ respectively; 
\begin{eqnarray}
&&  i T_{ij}(p_1,p_2,p_3,p_4)\nonumber\\
&=& \sum_{\alpha=1}^2\int\frac{d^D k}{(2\pi)^D}
    (-i h_i^\alpha) \left(\frac{1+\gamma_5}{2}\right) 
    \frac{i(k\!\!\!/+m_{N^\alpha}^{})}{k^2-m_{N^\alpha}^2}
     \left(\frac{1+\gamma_5}{2}\right) (-i h_j^\alpha)
    \frac{i}{(k+p_1)^2-m_{S^\pm}^2}\nonumber\\
&&     
\times(-i \sqrt{2} \kappa v)  \frac{i}{(k+p_1+p_3)^2-m_{\eta}^2}
    (-i \sqrt{2} \kappa v)
    \frac{i}{(k+p_1+p_3+p_4)^2-m_{S^\pm}^2} 
\end{eqnarray}

The Majorana mass matrix of the left handed neutrinos is generated
by connecting the external lines of $T_{ij}(p_1,p_2,p_3,p_4)$,
$\ell_R^i$ and $H^+$ with $\nu_L^i$ (also $\ell_R^i$, $H^+$ with
$\nu_L^i$) by the Yukawa coupling, and integrate all the internal
momenta taking into account the momentum conservation. There are two
Feynman diagrams which give exactly the same contribution, corresponding
to the way of connecting which $H^\pm$ couples to $\nu_L^i$:
see Fig.~\ref{diag-numass}. The mass matrix is calculated as 
\begin{eqnarray}
i M^2_{ij}
&=&+\int\frac{d^D p_1}{(2\pi)^D}\int\frac{d^D p_2}{(2\pi)^D}
    \frac{i}{p_1^2-m_{H^\pm}^2}
    (-i y_{\ell_i}^{\rm SM} \tan\beta)
    \left(\frac{1-\gamma_5}{2}\right)
    \frac{i (p_1\!\!\!\!/+m_{\ell^i}^{})}{p_1^2-m_{\ell^i}^2} \nonumber\\ 
&&\times  iT_{ij}(p_1,p_2,-p_1,-p_2)  \frac{i (p_2\!\!\!\!/+m_{\ell^j}^{})}{p_2^2-m_{\ell^j}^2} 
    \left(\frac{1-\gamma_5}{2}\right)
    (-i y_{\ell^j}^{\rm SM} \tan\beta)
    \frac{i}{p_2^2-m_{H^\pm}^2}
\nonumber \\
&&+\int\frac{d^D p_1}{(2\pi)^D}\int\frac{d^D p_2}{(2\pi)^D}
    \frac{i}{p_1^2-m_{H^\pm}^2}
    (-i y_{\ell_i}^{\rm SM} \tan\beta)
    \left(\frac{1-\gamma_5}{2}\right)
    \frac{i (p_1\!\!\!\!/+m_{\ell^i}^{})}{p_1^2-m_{\ell^i}^2}  \nonumber\\ 
&&\times iT_{ij}(p_1,p_2,-p_2,-p_1)
    \frac{i (p_2\!\!\!\!/+m_{\ell^j}^{})}{p_2^2-m_{\ell^j}^2} 
    \left(\frac{1-\gamma_5}{2}\right)
    (-i y_{\ell^j}^{\rm SM} \tan\beta)
    \frac{i}{p_2^2-m_{H^\pm}^2}\nonumber\\
 &=& + 4 \kappa^2 v^2 \tan^2\beta     \left(\frac{1-\gamma_5}{2}\right)
\sum_{\alpha=1}^2 \left[(y_{\ell_i}^{\rm SM}
  h_i^\alpha) 
      (y_{\ell^j}^{\rm SM} h_j^\alpha )  \left\{  \int\frac{d^D k}{(2\pi)^D}
    \frac{m_{N^\alpha}^{}}{k^2-m_{N^\alpha}^2}\frac{1}{k^2-m_{\eta}^2} \right.\right.
\nonumber\\ 
&&
\times 
 \left(
    \int\frac{d^D p_1}{(2\pi)^D}
    \frac{p_1\!\!\!\!/}{p_1^2-m_{\ell^i}^2}
    \frac{1}{p_1^2-m_{H^\pm}^2}
    \frac{1}{(k+p_1)^2-m_{S^\pm}^2}\right) \nonumber\\ 
&&\left.\left.
\times
\left(
\int\frac{d^D p_2}{(2\pi)^D}
    \frac{(-p_2\!\!\!\!/)}{(-p_2)^2-m_{\ell^j}^2}
\frac{1}{(-p_2)^2-m_{H^\pm}^2}
    \frac{1}{(k+(-p_2))^2-m_{S^\pm}^2} \right)\right\}\right],  
\end{eqnarray}
where we used momentum conservation law $p_3=-p_1$ and $p_4=-p_2$
($p_3=-p_4$ and $p_4=-p_1$) for the first (the second) diagram,
neglecting the invariant mass of neutrinos. 
By using Passarino-Veltman formalism for one-loop integral functions~\cite{passarino-veltman}, we
obtain the expression as 
\begin{eqnarray}
i M_{ij}
 &=& + \left(\frac{1}{16\pi^2}\right)^2
  \left(\frac{1-\gamma_5}{2}\right) 
\sum_{\alpha=1}^2 \left[
   \frac{4 \kappa^2 v^2 \tan^2\beta m_{N^\alpha}^{}}{m_{N^\alpha}^2-m_\eta^2} 
  \frac{(y_{\ell_i}^{\rm SM}
  h_i^\alpha) }{m_{H^\pm}^2-m_{\ell^i}^2}
  \frac{(y_{\ell_j}^{\rm SM}
  h_j^\alpha) }{m_{H^\pm}^2-m_{\ell^j}^2}\right.
\nonumber\\ 
&&
\times   \left\{  \int\frac{d^D k}{(2\pi)^D}
    \left( \frac{m_{N^\alpha}^2}{k^2-m_{N^\alpha}^2}-\frac{m_\eta^2}{k^2-m_{\eta}^2} 
      \right) 
 \left(
B_1(k^2,m_{H^\pm}^2,m_{S^\pm}^2)-B_1(k^2, m_{\ell^i}^2, m_{S^\pm}^2)
\right) \right. \nonumber\\
 &&
  \left.\left. \times
 \left(
B_1(k^2,m_{H^\pm}^2,m_{S^\pm}^2)-B_1(k^2, m_{\ell^j}^2, m_{S^\pm}^2)
\right)
  \frac{}{}        \right\}\right],
\end{eqnarray}
where
$B_1(k^2,m_1,m_2)$ is the tensor coefficient by Passarino and Veltman~\cite{passarino-veltman}.
As $m_{\ell}^2 \ll m_{H^\pm}^2$, we neglect $m_{\ell}^2$ in the
expression and obtain
\begin{eqnarray}
i M_{ij} &=& + \left(\frac{1}{16\pi^2}\right)^2
  \left(\frac{1-\gamma_5}{2}\right)
  \frac{4 \kappa^2 v^2 \tan^2\beta}{m_{H^\pm}^4} 
\sum_{\alpha=1}^2  \left[
   \frac{m_{N^\alpha}^{}}{m_{N^\alpha}^2-m_\eta^2} 
  (y_{\ell_i}^{\rm SM} h_i^\alpha) (y_{\ell_j}^{\rm SM} h_j^\alpha)
\right. 
\nonumber\\ 
&&
 \left. \hspace*{-2cm}
\times     \int\frac{d^D k}{(2\pi)^D}
 \left(
B_1(k^2,m_{H^\pm}^2,m_{S^\pm}^2)-B_1(k^2, m_{\ell^i}^2, m_{S^\pm}^2)
\right)^2    \left( \frac{m_{N^\alpha}^2}{k^2-m_{N^\alpha}^2}-\frac{m_\eta^2}{k^2-m_{\eta}^2} 
      \right) 
         \right].
\end{eqnarray}
As the asymptotic behavior $k^2 \to \infty$ of the $B_1$ function is 
\begin{eqnarray}
   \left\{B_1(k^2,m_H^{},m_S^{})-B_1(k^2,0,m_S^{})\right\} \sim 1/k^2,
\end{eqnarray}
so that the integral over $d^4 k$ is not divergent. For numerical
evaluations, we work in the Euclideanized momentum space, 
\begin{eqnarray}
  k^2 = - k_E^2,  {\hspace{1cm}}   d^4 k = i d^4 k_E^{} =i \pi^2 k_E^2 d^4(k_E^2). 
\end{eqnarray}
and introduce the cutoff scale $\Lambda$ which is a very
large number as compared to the scale of $m_H^{}$ or $m_S^{}$ and so on.
Then we obtain the expression in Eq.~(\ref{eq:mij}).

\subsection{Line photon flux from DM annihilation}

The differential flux of gamma-ray from a DM annihilation 
near the center of our galaxy is given as
\begin{eqnarray}
\frac{d \Phi}{d \Omega}(E,\psi)
 = \frac{1}{4 \pi m_{\eta}^2}
 \left[\langle\sigma v(\rightarrow\gamma\gamma)\rangle\delta(E-m_{\eta})+\langle\sigma v(\rightarrow f\bar{f})\rangle\frac{d N}{d E}\right]
 \int_{l.o.s} dl(\psi) \rho(l)^2,
 \label{DeffFlux}
\end{eqnarray}
where $\psi$ is the angle to the galactic center direction, 
$\rho(l)$ is the mass density distribution for the DM 
and we will integrate it along the line of sight $l$~\cite{Bertone:2004pz}.
The first term in the right hand side of Eq.~(\ref{DeffFlux}) denotes 
 the line spectrum comes from the annihilation into two photons,
 and the second term does to the continuous one.
The latter with the differential photon spectrum $d N/dE$ dominantly comes from the decay of pions produced by the fragmentation or decay
 of DM annihilation final state $f$ such as b-quark or $\tau$ lepton.

For the integration around the line of sight axis 
over the solid angle $\Delta\Omega$, we introduce 
a dimensionless function~\cite{Bergstrom:1997fj}
\begin{eqnarray}
J(\psi) = \frac{1}{8.5 {\rm kpc}}\left(\frac{1}{0.3 {\rm GeV cm^{-3}}}\right)\int_{l.o.s} dl(\psi) \rho(l)^2,
\label{J}
\end{eqnarray}
in which all informations about the DM halo model are encoded.
For a given $\Delta\Omega\ \sim 10^{-3} (10^{-5})$,
one may find, for instance, $J \sim 10^3 (10^4)$ for NFW density profile~\cite{Navarro:1996gj}
 and $J \sim 10 (10)$ for isothermal~\cite{Cesarini:2003nr}.
With the integration over angle and substitution of Eq.~(\ref{J}), we obtain Eq.~(\ref{LineGammaFlux}).

\subsection{High Temperature Expansion}

Let us calculate the integral $I_f$ by using the 
high temperature expansion~\cite{hte}. For the case of bosonic contrbutions, we have
\begin{eqnarray}
   I_B(a) &=& \int_0^\infty dx \, x^2 \log[1 - e^{-\sqrt{x^2+a^2}}] \nonumber \\
          &=& - \frac{2 \pi^2}{T^4} 
  \left\{ - \frac{1}{\beta} \int_0^\infty \frac{d^3 p}{(2 \pi)^3} 
               \log [1 - e^{-\beta \omega}]  \right\},
\end{eqnarray}
where $\omega=\sqrt{p^2 + m^2}$. The integral is expanded as 
\begin{eqnarray}
&&- \frac{1}{\beta} \int_0^\infty \frac{d^3 p}{(2 \pi)^3} 
               \log [1 - e^{-\beta \omega}] \nonumber \\
&& = \frac{\pi^2}{90} T^4 - \frac{1}{24} m^2 T^2 + \frac{1}{12 \pi} m^3 T
   + \frac{m^4}{64\pi^2} 
  \left[ \log\left(\frac{m^2}{16 \pi^2 T^2}\right) 
        + 2 \gamma_E^{} - \frac{3}{2} \right] + ... ,
\end{eqnarray}
where $\gamma_E^{} = 0.5772$ is the Euler constant.
For fermions, we obtain 
\begin{eqnarray}
   I_F(a) &=& \int_0^\infty dx \, x^2 \log[1 + e^{-\sqrt{x^2+a^2}}] \nonumber \\
          &=& - \frac{2 \pi^2}{T^4} 
  \left\{ - \frac{1}{\beta} \int_0^\infty \frac{d^3 p}{(2 \pi)^3} 
               \log [1 + e^{-\beta \omega}]  \right\}\nonumber  \\
          &=& - \frac{2 \pi^2}{T^4} 
  \left\{ 
\frac{7 \pi^2}{720} T^4 - \frac{1}{48} m^2 T^2  
   - \frac{m^4}{64 \pi^2} 
  \left[ \log\left(\frac{m^2}{\pi^2 T^2}\right) 
        + 2 \gamma_E^{} - \frac{3}{2} \right] + ... 
 \right\}.
\end{eqnarray}


\newpage

\vspace*{-4mm}
\begin{thebibliography}{1}

\bibitem{aks-prl} % AKS PRL
%\cite{Aoki:2008av}
%\bibitem{Aoki:2008av}
  M.~Aoki, S.~Kanemura and O.~Seto,
  %``Neutrino mass, Dark Matter and Baryon Asymmetry via TeV-Scale Physics
  %without Fine-Tuning,''
  Phys.\ Rev.\ Lett.\  {\bf 102}, 051805 (2009).
%  [arXiv:0807.0361 [hep-ph]].
  %%CITATION = PRLTA,102,051805;%%
 
\bibitem{lep-data}

%\cite{Schwetz:2008er}
% \bibitem{Schwetz:2008er}
 T.~Schwetz, M.~Tortola and J.~W.~F.~Valle,
 %``Three-flavour neutrino oscillation update,''
 New J.\ Phys.\  {\bf 10} (2008) 113011.
% [arXiv:0808.2016 [hep-ph]].

 \bibitem{wimp}
%
%\cite{Komatsu:2008hk}
%\bibitem{Komatsu:2008hk}
 E.~Komatsu {\it et al.}  [WMAP Collaboration],
 %``Five-Year Wilkinson Microwave Anisotropy Probe (WMAP\altaffilmark 1 )
 %Observations:Cosmological Interpretation,''
 Astrophys.\ J.\ Suppl.\  {\bf 180} (2009) 330
%  [arXiv:0803.0547 [astro-ph]].
 %%CITATION = APJSA,180,330;%%

 \bibitem{sakharov}
%\SakharovDJ
%\lref\SakharovDJ{
  A.~D.~Sakharov,
  %``Violation of CP Invariance, c Asymmetry, and Baryon Asymmetry of the
  %Universe,''
  Pisma Zh.\ Eksp.\ Teor.\ Fiz.\  {\bf 5}, 32 (1967).
%  [JETP Lett.\  {\bf 5}, 24 (1967\ SOPUA,34,392-393.1991\ UFNAA,161,61-64.1991)].
  %%CITATION = UFNAA,161,NO.561;%%
%}
 
 \bibitem{see-saw}
         T.~Yanagida, In Proceedings of Workshop on  {\it the Unified 
        Theory and the Baryon Number in the Universe}, p.95 KEK Tsukuba,
        Japan (1979);

         M. Gell-Mann, P.~Ramond and R.~Slansky, 
        in Proceedings of Workshop {\it Supergravity}, p.315, Stony
        Brook, New York, 1979.
  
  
 \bibitem{FY}
         M.~Fukugita and T.~Yanagida,
  %``Baryogenesis Without Grand Unification,''
  Phys.\ Lett.\  B {\bf 174}, 45 (1986).


\bibitem{Jungman:1995df}
 For a review, e.g., G.~Jungman, M.~Kamionkowski and K.~Griest,
  %``Supersymmetric dark matter,''
  Phys.\ Rept.\  {\bf 267}, 195 (1996);
        
  C.~Munoz,
  %``Dark matter detection in the light of recent experimental results,''
  Int.\ J.\ Mod.\ Phys.\  A {\bf 19}, 3093 (2004).
%  [arXiv:hep-ph/0309346].

\bibitem{dec-theorem}
%\AppelquistTG
%\lref\AppelquistTG{
  T.~Appelquist and J.~Carazzone,
  %``Infrared Singularities And Massive Fields     ,''
  Phys.\ Rev.\  D {\bf 11}, 2856 (1975).
  %%CITATION = PHRVA,D11,2856;%%
%}

\bibitem{zee}
%\ZeeAI
%\lref\ZeeAI{
 A.~Zee,
 %``A Theory Of Lepton Number Violation, Neutrino Majorana Mass, And
 %Oscillation,''
 Phys.\ Lett.\  B {\bf 93}, 389 (1980)
 [Erratum-ibid.\  B {\bf 95}, 461 (1980)];
 %%CITATION = PHLTA,B93,389;%%

        %\ZeeRJ
%\lref\ZeeRJ{

        A.~Zee,
 %``Charged Scalar Field And Quantum Number Violations,''
 Phys.\ Lett.\  B {\bf 161}, 141 (1985).
 %%CITATION = PHLTA,B161,141;%%

 \bibitem{zee-ph2}
%\cite{Petcov:1982en}
%\bibitem{Petcov:1982en}
 S.~T.~Petcov,
 %``Remarks On The Zee Model Of Neutrino Mixing (Mu $\to$ E Gamma, Heavy
 %Neutrino $\to$ Light Neutrino Gamma, Etc.),''
 Phys.\ Lett.\  B {\bf 115} (1982) 401.
 %%CITATION = PHLTA,B115,401;%%
 
 \bibitem{zee-ph}
%\KanemuraBQ
%\lref\KanemuraBQ{
 S.~Kanemura, T.~Kasai, G.~L.~Lin, Y.~Okada, J.~J.~Tseng and C.~P.~Yuan,
 %``Phenomenology of Higgs bosons in the Zee-model,''
 Phys.\ Rev.\  D {\bf 64}, 053007 (2001).
%  [arXiv:hep-ph/0011357].
 %%CITATION = PHRVA,D64,053007;%%
%}

 \bibitem{zee-ex1}

%\cite{Jarlskog:1998uf}
%\bibitem{Jarlskog:1998uf}
 % C.~Jarlskog, M.~Matsuda, S.~Skadhauge and M.~Tanimoto,
 %``Zee mass matrix and bi-maximal neutrino mixing,''
 % Phys.\ Lett.\  B {\bf 449} (1999) 240
 %[arXiv:hep-ph/9812282].
 %%CITATION = PHLTA,B449,240;%%
         
 %\cite{Frampton:2001eu}
%\bibitem{Frampton:2001eu}
 P.~H.~Frampton, M.~C.~Oh and T.~Yoshikawa,
 %``Zee model confronts SNO data,''
 Phys.\ Rev.\  D {\bf 65} (2002) 073014
 [arXiv:hep-ph/0110300];
 %%CITATION = PHRVA,D65,073014;%%

%\cite{Kitabayashi:2001it}
%\bibitem{Kitabayashi:2001it}
 T.~Kitabayashi and M.~Yasue,
 %``Large solar neutrino mixing in an extended Zee model,''
 Int.\ J.\ Mod.\ Phys.\  A {\bf 17} (2002) 2519;
% [arXiv:hep-ph/0112287];
 %%CITATION = IMPAE,A17,2519;%%

%\cite{Hasegawa:2003by}
%\bibitem{Hasegawa:2003by}
 K.~Hasegawa, C.~S.~Lim and K.~Ogure,
 %``Escape from washing out of baryon number in a two-zero-texture general  Zee
 %model compatible with the LMA-MSW solution,''
 Phys.\ Rev.\  D {\bf 68} (2003) 053006;
% [arXiv:hep-ph/0303252];
 %%CITATION = PHRVA,D68,053006;%%

 %\cite{Balaji:2001ex}
%\bibitem{Balaji:2001ex}
 K.~R.~S.~Balaji, W.~Grimus and T.~Schwetz,
 %``The solar LMA neutrino oscillation solution in the Zee model,''
 Phys.\ Lett.\  B {\bf 508} (2001) 301.
% [arXiv:hep-ph/0104035].
 %%CITATION = PHLTA,B508,301;%%

 \bibitem{ZeeBabu}

         %\cite{Zee:1985id}
%\bibitem{Zee:1985id}
 A.~Zee,
 %``Quantum Numbers Of Majorana Neutrino Masses,''
 Nucl.\ Phys.\ B {\bf 264}, 99 (1986);
 %%CITATION = NUPHA,B264,99;%%

%\cite{Babu:1988ki}
%\bibitem{Babu:1988ki}
 K.~S.~Babu,
 %``MODEL OF 'CALCULABLE' MAJORANA NEUTRINO MASSES,''
 Phys.\ Lett.\ B {\bf 203}, 132 (1988).
 %%CITATION = PHLTA,B203,132;%%

\bibitem{ZeeBabu-ph}
%\cite{Babu:2002uu}
%\bibitem{Babu:2002uu}
  K.~S.~Babu and C.~Macesanu,
 %``Two-loop neutrino mass generation and its experimental consequences,''
  Phys.\ Rev.\ D {\bf 67}, 073010 (2003);
%  [arXiv:hep-ph/0212058];
 %%CITATION = HEP-PH 0212058;%%
%\cite{Sierra:2006gb}
%\bibitem{Sierra:2006gb}
  D.~A.~Sierra and M.~Hirsch,
%   ``Experimental tests for the babu-zee two-loop model of Majorana neutrino
 %masses,''
 arXiv:hep-ph/0609307.
 %%CITATION = HEP-PH 0609307;%%

 \bibitem{Zee-LG}
 %\cite{Sahu:2008aw}
%\bibitem{Sahu:2008aw}
 N.~Sahu and U.~Sarkar,
 %``Extended Zee model for Neutrino Mass, Leptogenesis and Sterile Neutrino
 %like Dark Matter,''
 Phys.\ Rev.\  D {\bf 78} (2008) 115013.
% [arXiv:0804.2072 [hep-ph]].
 %%CITATION = PHRVA,D78,115013;%%


\bibitem{ZeeBabu-LG}
%\cite{Chen:2008ma}
%\bibitem{Chen:2008ma}
 C.~S.~Chen, C.~Q.~Geng and D.~V.~Zhuridov,
 %``Leptogenesis in the minimal extnsion of the Babu-Zee model,''
 arXiv:0806.2698 [hep-ph].
 %%CITATION = ARXIV:0806.2698;%%

 \bibitem{knt}
 %\KraussPX
%  \lref\KraussPX{
 L.~M.~Krauss, S.~Nasri and M.~Trodden,
 %``A model for neutrino masses and dark matter,''
 Phys.\ Rev.\  D {\bf 67}, 085002 (2003).
%  [arXiv:hep-ph/0011357].
 %%CITATION = PHRVA,D67,085002;%%

 \bibitem{kingman-seto}
%\CheungXM
        %\lref\CheungXM{
 K.~Cheung and O.~Seto,
 %``Phenomenology of TeV right-handed neutrino and the dark matter model,''
 Phys.\ Rev.\  D {\bf 69}, 113009 (2004).
%  [arXiv:hep-ph/0403003].
 %%CITATION = PHRVA,D69,113009;%%
%}

 \bibitem{NRandDM}
%\cite{Ma:2006km}
%\bibitem{Ma:2006km}
 E.~Ma,
 %``Verifiable radiative seesaw mechanism of neutrino mass and dark matter,''
 Phys.\ Rev.\  D {\bf 73}, 077301 (2006);
%  [arXiv:hep-ph/0601225];
 %%CITATION = PHRVA,D73,077301;%%
%}

%\KuboYX
%\lref\KuboYX{
 J.~Kubo, E.~Ma and D.~Suematsu,
 %``Cold dark matter, radiative neutrino mass, mu --> e gamma, and
 %neutrinoless double beta decay,''
 Phys.\ Lett.\  B {\bf 642}, 18 (2006);
%  [arXiv:hep-ph/0604114];
 %%CITATION = PHLTA,B642,18;%%
%}

%\cite{Ma:2007gq}
%\bibitem{Ma:2007gq}
 E.~Ma,
 %``Z_3 Dark Matter and Two-Loop Neutrino Mass,''
 Phys.\ Lett.\  B {\bf 662} (2008) 49;
% [arXiv:0708.3371 [hep-ph]];
 %%CITATION = PHLTA,B662,49;%%

% %\cite{Porto:2008hb}
%\bibitem{Porto:2008hb}
 R.~A.~Porto and A.~Zee,
 %``Neutrino Mixing and the Private Higgs,''
 Phys.\ Rev.\  D {\bf 79} (2009) 013003; 
% [arXiv:0807.0612 [hep-ph]];
 %%CITATION = PHRVA,D79,013003;%%

%\cite{Suematsu:2009ww}
%\bibitem{Suematsu:2009ww}
 D.~Suematsu, T.~Toma and T.~Yoshida,
 %``Reconciliation of CDM abundance and $\mu\to e\gamma$ in a radiative seesaw
 %model,''
 arXiv:0903.0287 [hep-ph].
 %%CITATION = ARXIV:0903.0287;%%
         
 \bibitem{NRandLG}
 %\cite{Ma:2006fn}
%\bibitem{Ma:2006fn}
 E.~Ma,
 %``Common origin of neutrino mass, dark matter, and baryogenesis,''
 Mod.\ Phys.\ Lett.\  A {\bf 21} (2006) 1777;
% [arXiv:hep-ph/0605180]; 
 %%CITATION = MPLAE,A21,1777;%%

%\HambyeZN
%\lref\HambyeZN{
% T.~Hambye, et al., %K.~Kannike, E.~Ma and M.~Raidal,
 T.~Hambye, K.~Kannike, E.~Ma and M.~Raidal,
 %``Emanations of dark matter: Muon anomalous magnetic moment, radiative
 %neutrino mass, and novel leptogenesis at the TeV scale,''
 Phys.\ Rev.\  D {\bf 75}, 095003 (2007).
%  [arXiv:hep-ph/0609228];
 %%CITATION = PHRVA,D75,095003;%%

\bibitem{NRandEWBG}
%\cite{Babu:2007sm}
%\bibitem{Babu:2007sm}
 K.~S.~Babu and E.~Ma,
 %``Singlet fermion dark matter and electroweak baryogenesis with radiative
 %neutrino mass,''
 Int.\ J.\ Mod.\ Phys.\  A {\bf 23} (2008) 1813.
% [arXiv:0708.3790 [hep-ph]].
 %%CITATION = IMPAE,A23,1813;%%

\bibitem{ewbg-thdm2}
%\KanemuraCH
%\lref\KanemuraCH{
  S.~Kanemura, Y.~Okada and E.~Senaha,
  %``Electroweak baryogenesis and quantum corrections to the triple Higgs  boson
  %coupling,''
  Phys.\ Lett.\  B {\bf 606}, 361 (2005).
%  [arXiv:hep-ph/0411354].
  %%CITATION = PHLTA,B606,361;%%
%}

 \bibitem{ewbg-thdm}

%\cite{Bochkarev:1990fx}
%\bibitem{Bochkarev:1990fx}
  A.~I.~Bochkarev, S.~V.~Kuzmin and M.~E.~Shaposhnikov,
  %``Electroweak baryogenesis and the Higgs boson mass problem,''
  Phys.\ Lett.\  B {\bf 244} (1990) 275;
  %%CITATION = PHLTA,B244,275;%%

%\cite{Nelson:1991ab}
%\bibitem{Nelson:1991ab}
  A.~E.~Nelson, D.~B.~Kaplan and A.~G.~Cohen,
  %``Why there is something rather than nothing: Matter from weak
  %interactions,''
  Nucl.\ Phys.\  B {\bf 373}, 453 (1992);
  %%CITATION = NUPHA,B373,453;%%

        %\cite{Turok:1991uc}
%\bibitem{Turok:1991uc}
  N.~Turok and J.~Zadrozny,
  %``Phase transitions in the two doublet model,''
  Nucl.\ Phys.\  B {\bf 369}, 729 (1992);
  %%CITATION = NUPHA,B369,729;%%

         
%\cite{Funakubo:1993jg}
%\bibitem{Funakubo:1993jg}
  K.~Funakubo, A.~Kakuto and K.~Takenaga,
  %``The Effective potential of electroweak theory with two massless Higgs
  %doublets at finite temperature,''
  Prog.\ Theor.\ Phys.\  {\bf 91}, 341 (1994);
%  [arXiv:hep-ph/9310267];
  %%CITATION = PTPKA,91,341;%%
         
%\cite{Davies:1994id}
%\bibitem{Davies:1994id}
        A.~T.~Davies, C.~D.~froggatt, G.~Jenkins and R.~G.~Moorhouse,
  %``Baryogenesis constraints on two Higgs doublet models,''
  Phys.\ Lett.\  B {\bf 336}, 464 (1994);
  %%CITATION = PHLTA,B336,464;%%

%\ClineDG
%\lref\ClineDG{
  J.~M.~Cline, K.~Kainulainen and A.~P.~Vischer,
  %``Dynamics of two Higgs doublet CP violation and baryogenesis at the
  %electroweak phase transition,''
  Phys.\ Rev.\  D {\bf 54}, 2451 (1996);
%  [arXiv:hep-ph/9506284];
  %%CITATION = PHRVA,D54,2451;%%
%}

 \bibitem{ewbg-thdm3}
         
         %\FrommeCM
%\lref\FrommeCM{
  L.~Fromme, S.~J.~Huber and M.~Seniuch,
  %``Baryogenesis in the two-Higgs doublet model,''
  JHEP {\bf 0611}, 038 (2006).
%  [arXiv:hep-ph/0605242];
  %%CITATION = JHEPA,0611,038;%%
%}

\bibitem{glashow-weinberg}

%\GlashowNT
%\lref\GlashowNT{
  S.~L.~Glashow and S.~Weinberg,
  %``Natural Conservation Laws For Neutral Currents,''
  Phys.\ Rev.\  D {\bf 15}, 1958 (1977).
  %%CITATION = PHRVA,D15,1958;%%
%}
        
\bibitem{barger}
  V.~D.~Barger, J.~L.~Hewett and R.~J.~N.~Phillips,
  %``NEW CONSTRAINTS ON THE CHARGED HIGGS SECTOR IN TWO HIGGS DOUBLET MODELS,''
  Phys.\ Rev.\  D {\bf 41}, 3421 (1990).
  %%CITATION = PHRVA,D41,3421;%%

 \bibitem{grossman}

%\GrossmanJB
%\lref\GrossmanJB{
  Y.~Grossman,
  %``Phenomenology of models with more than two Higgs doublets,''
  Nucl.\ Phys.\  B {\bf 426}, 355 (1994).
%  [arXiv:hep-ph/9401311].
  %%CITATION = NUPHA,B426,355;%%
%}
        
\bibitem{typeX}

%\cite{Aoki:2009ha}
%\bibitem{Aoki:2009ha}
  M.~Aoki, S.~Kanemura, K.~Tsumura and K.~Yagyu,
  %``Models of Yukawa interaction in the two Higgs doublet model, and their
  %collider phenomenology,''
  arXiv:0902.4665 [hep-ph].        
%%CITATION = ARXIV:0902.4665;%%

\bibitem{typeX2}
        
        
%\cite{Su:2009fz}
%\bibitem{Su:2009fz}
  S.~Su and B.~Thomas,
  %``The LHC Discovery Potential of a Leptophilic Higgs,''
  arXiv:0903.0667 [hep-ph]; 
  %%CITATION = ARXIV:0903.0667;%%
        
%\cite{Logan:2009uf}
%\bibitem{Logan:2009uf}
  H.~E.~Logan and D.~MacLennan,
  %``Charged Higgs phenomenology in the lepton-specific two Higgs doublet
  %model,''
  arXiv:0903.2246 [hep-ph].
  %%CITATION = ARXIV:0903.2246;%%

 \bibitem{hhg}
%\GunionWE
%\lref\GunionWE{
  J.~F.~Gunion, H.~E.~Haber, G.~L.~Kane and S.~Dawson,
%  J.~F.~Gunion, et al., 
  ``{\it The Higgs Hunters's Guide}'' (Addison Wesley, 1990).
  %%CITATION = BNL-41644;%%
%}

\bibitem{pu-thdm}
%}

        %\cite{Huffel:1980sk}
%\bibitem{Huffel:1980sk}
  H.~Huffel and G.~Pocsik,
  %``Unitarity Bounds On Higgs Boson Masses In The Weinberg-Salam Model With Two
  %Higgs Doublets,''
  Z.\ Phys.\  C {\bf 8}, 13 (1981);
  %%CITATION = ZEPYA,C8,13;%%


    %\cite{Maalampi:1991fb}
%\bibitem{Maalampi:1991fb}
  J.~Maalampi, J.~Sirkka and I.~Vilja,
  %``Tree Level Unitarity And Triviality Bounds For Two Higgs Models,''
  Phys.\ Lett.\  B {\bf 265}, 371 (1991);
  %%CITATION = PHLTA,B265,371;%%
        
        %\KanemuraHM
%\lref\KanemuraHM{
  S.~Kanemura, T.~Kubota and E.~Takasugi,
  %``Lee-Quigg-Thacker bounds for Higgs boson masses in a two doublet model,''
  Phys.\ Lett.\  B {\bf 313}, 155 (1993); 
%  [arXiv:hep-ph/9303263].
  %%CITATION = PHLTA,B313,155;%%
%}
        
%\AkeroydWC
%\lref\AkeroydWC{
  A.~G.~Akeroyd, A.~Arhrib and E.~M.~Naimi,
  %``Note on tree-level unitarity in the general two Higgs doublet model,''
  Phys.\ Lett.\  B {\bf 490}, 119 (2000).
%  [arXiv:hep-ph/0006035];
  %%CITATION = PHLTA,B490,119;%%
%}

    %\cite{Ginzburg:2003fe}
%\bibitem{Ginzburg:2003fe}
  I.~F.~Ginzburg and I.~P.~Ivanov,
  %``Tree-level unitarity constraints in the 2HDM with CP-violation,''
  arXiv:hep-ph/0312374.
  %%CITATION = HEP-PH/0312374;%%

\bibitem{rge-thdm}

        %\NieYN
%\lref\NieYN{
  S.~Nie and M.~Sher,
  %``Vacuum stability bounds in the two-Higgs doublet model,''
  Phys.\ Lett.\  B {\bf 449}, 89 (1999);
%  [arXiv:hep-ph/9811234];
  %%CITATION = PHLTA,B449,89;%%
%}
        
%\KanemuraXF
%\lref\KanemuraXF{
  S.~Kanemura, T.~Kasai and Y.~Okada,
  %``Mass bounds of the lightest CP-even Higgs boson in the  two-Higgs-doublet
  %model,''
  Phys.\ Lett.\  B {\bf 471}, 182 (1999).
%  [arXiv:hep-ph/9903289].
  %%CITATION = PHLTA,B471,182;%%

\bibitem{passarino-veltman}
%\PassarinoJH
%\lref\PassarinoJH{
  G.~Passarino and M.~J.~G.~Veltman,
  %``One Loop Corrections For E+ E- Annihilation Into Mu+ Mu- In The Weinberg
  %Model,''
  Nucl.\ Phys.\  B {\bf 160}, 151 (1979).
  %%CITATION = NUPHA,B160,151;%%
%}

 \bibitem{mns}
%\MakiMU
%\lref\MakiMU{
  Z.~Maki, M.~Nakagawa and S.~Sakata,
  %``Remarks on the unified model of elementary particles,''
  Prog.\ Theor.\ Phys.\  {\bf 28}, 870 (1962).
  %%CITATION = PTPKA,28,870;%%
%}

%\cite{Brooks:1999pu}
\bibitem{MEGA}
   M.~L.~Brooks {\it et al.}  [MEGA Collaboration],
   %``New Limit for the Family-Number Non-conserving Decay mu+ to e +_gamma,''
   Phys.\ Rev.\ Lett.\  {\bf 83} (1999) 1521
%  [arXiv:hep-ex/9905013].
   %%CITATION = PRLTA,83,1521;%%

 \bibitem{bsgamma}

%\cite{Barberio:2008fa}
%\bibitem{Barberio:2008fa}
 E.~Barberio {\it et al.}  [Heavy Flavor Averaging Group],
 %``Averages of $b-$hadron and $c-$hadron Properties at the End of 2007,''
 arXiv:0808.1297 [hep-ex].
 %%CITATION = ARXIV:0808.1297;%%

 \bibitem{comment1}

The unitarity bound for the scattering process $h \eta \to H^+S^-$ turns
         out to give $\kappa \lsim 25$, whereas $\tan\beta$ should not be
         too large for successful electroweak baryogenesis\cite{ewbg-thdm,ewbg-thdm3}.

\bibitem{john}
  J.~McDonald, Phys.\ Rev.\  D {\bf 50}, 3637 (1994); 

        For a resent study, see e.g., 
  H.~Sung Cheon, S.~K.~Kang and C.~S.~Kim, 
  J. Cosmol. Astropart. Phys. 05 (2008) 004.

\bibitem{kt}
                E.~W.~Kolb and M.~S.~Turner, {\it The Early Universe}
                (Addison-Wesley, 1990).

\bibitem{CDMS}
  D.~S.~Akerib {\it et al.}  [CDMS Collaboration],
  %``First results from the cryogenic dark matter search in the Soudan
  %Underground Lab,''
  Phys.\ Rev.\ Lett.\  {\bf 93}, 211301 (2004); 
  Phys.\ Rev.\ Lett.\ {\bf 96}, 011302 (2006).
%  [arXiv:astro-ph/0405033].

\bibitem{Ellis:2008hf}
  J.~R.~Ellis, K.~A.~Olive and C.~Savage,
  %``Hadronic Uncertainties in the Elastic Scattering of Supersymmetric Dark
  %Matter,''
  Phys.\ Rev.\  D {\bf 77}, 065026 (2008).

\bibitem{xenon}
  J.~Angle {\it et al.}  [XENON Collaboration],
  %``First Results from the XENON10 Dark Matter Experiment at the Gran Sasso
  %National Laboratory,''
  Phys.\ Rev.\ Lett.\  {\bf 100}, 021303 (2008).
%  [arXiv:0706.0039 [astro-ph]].

\bibitem{Cerdeno}
  D.~G.~Cerdeno, C.~Munoz and O.~Seto,
  %``Right-handed sneutrino as thermal dark matter,''
  Phys.\ Rev.\  D {\bf 79}, 023510 (2009);

        D.~G.~Cerdeno and O.~Seto,
  %``Right-handed sneutrino dark matter in the NMSSM,''
  arXiv:0903.4677 [hep-ph].

\bibitem{pamela}
  O.~Adriani {\it et al.}  [PAMELA Collaboration],
  %``An anomalous positron abundance in cosmic rays with energies 1.5.100 GeV,''
  Nature {\bf 458}, 607 (2009).
%  [arXiv:0810.4995 [astro-ph]].

\bibitem{atic}
  J.~Chang {\it et al.},
  %``An Excess Of Cosmic Ray Electrons At Energies Of 300.800 Gev,''
  Nature {\bf 456}, 362 (2008).

\bibitem{annihilation}
For incomplete list of references, see e.g.,

        L.~Bergstrom, T.~Bringmann and J.~Edsjo,
  %``New Positron Spectral Features from Supersymmetric Dark Matter - a Way to
  %Explain the PAMELA Data?,''
  Phys.\ Rev.\  D {\bf 78}, 103520 (2008);

        M.~Cirelli and A.~Strumia,
  %``Minimal Dark Matter predictions and the PAMELA positron excess,''
  arXiv:0808.3867 [astro-ph];

        M.~Cirelli, M.~Kadastik, M.~Raidal and A.~Strumia,
  %``Model-independent implications of the e+, e-, anti-proton cosmic ray
  %spectra on properties of Dark Matter,''
  Nucl.\ Phys.\  B {\bf 813}, 1 (2009);

        V.~Barger, W.~Y.~Keung, D.~Marfatia and G.~Shaughnessy,
  %``PAMELA and dark matter,''
  Phys.\ Lett.\  B {\bf 672}, 141 (2009);
        
  J.~Hisano, M.~Kawasaki, K.~Kohri and K.~Nakayama,
  %``Positron/Gamma-Ray Signatures of Dark Matter Annihilation and Big-Bang
  %Nucleosynthesis,''
  Phys.\ Rev.\  D {\bf 79}, 063514 (2009);  

        J.~H.~Huh, J.~E.~Kim and B.~Kyae,
  %``Two dark matter components in N_{DM}MSSM and PAMELA data,''
  arXiv:0809.2601 [hep-ph].

\bibitem{decay}
For incomplete list of references, see e.g.,
        
  P.~f.~Yin, Q.~Yuan, J.~Liu, J.~Zhang, X.~j.~Bi and S.~h.~Zhu,
  %``PAMELA data and leptonically decaying dark matter,''
  Phys.\ Rev.\  D {\bf 79}, 023512 (2009);

        K.~Ishiwata, S.~Matsumoto and T.~Moroi,
  %``Cosmic-Ray Positron from Superparticle Dark Matter and the PAMELA
  %Anomaly,''
  arXiv:0811.0250 [hep-ph];  

        A.~Ibarra and D.~Tran,
  %``Decaying Dark Matter and the PAMELA Anomaly,''
  JCAP {\bf 0902}, 021 (2009).

\bibitem{AKSpamela}
  M.~Aoki, S.~Kanemura and O.~Seto, work in progress.

\bibitem{kingman}
%\cite{Cheung:2009si}
%\bibitem{Cheung:2009si}
  K.~Cheung, P.~Y.~Tseng and T.~C.~Yuan,
  %``Double-action dark matter, PAMELA and ATIC,''
  Phys.\ Lett.\  B {\bf 678}, 293 (2009)
  [arXiv:0902.4035 [hep-ph]].
  %%CITATION = PHLTA,B678,293;%%



\bibitem{dama}
  R.~Bernabei {\it et al.}  [DAMA Collaboration],
  %``First results from DAMA/LIBRA and the combined results with DAMA/NaI,''
  Eur.\ Phys.\ J.\  C {\bf 56}, 333 (2008).

%\cite{inelastic}
\bibitem{inelastic}
  D.~Tucker-Smith and N.~Weiner,
  %``Inelastic dark matter,''
  Phys.\ Rev.\  D {\bf 64}, 043502 (2001).
%  [arXiv:hep-ph/0101138]
        
\bibitem{lightWIMP}  
  e.g.,
  F.~Petriello and K.~M.~Zurek,
  %``DAMA and WIMP dark matter,''
  JHEP {\bf 0809}, 047 (2008);

        %  [arXiv:0806.3989 [hep-ph]].
  S.~Chang, A.~Pierce and N.~Weiner,
  %``Using the Energy Spectrum at DAMA/LIBRA to Probe Light Dark Matter,''
  arXiv:0808.0196 [hep-ph];

        C.~Savage, G.~Gelmini, P.~Gondolo and K.~Freese,
 %``Compatibility of DAMA/LIBRA dark matter detection with other searches,''
 JCAP {\bf 0904} (2009) 010;%%
        
  S.~Andreas, T.~Hambye and M.~H.~G.~Tytgat,
  %``WIMP dark matter, Higgs exchange and DAMA,''
  JCAP {\bf 0810}, 034 (2008).
%  [arXiv:0808.0255 [hep-ph]].

\bibitem{twostage}
     %\cite{Land:1992sm}
%\bibitem{Land:1992sm}
  D.~Land and E.~D.~Carlson,
  %``Two stage phase transition in two Higgs models,''
  Phys.\ Lett.\  B {\bf 292}, 107 (1992).
%  [arXiv:hep-ph/9208227].
  %%CITATION = PHLTA,B292,107;%%

\bibitem{hte}

%\AndersonZB
%\lref\AndersonZB{
  G.~W.~Anderson and L.~J.~Hall,
  %``The Electroweak Phase Transition And Baryogenesis,''
  Phys.\ Rev.\  D {\bf 45}, 2685 (1992);
  %%CITATION = PHRVA,D45,2685;%%
%}

        %\DineWR
%\lref\DineWR{
  M.~Dine, et al., %R.~G.~Leigh, P.~Y.~Huet, A.~D.~Linde and D.~A.~Linde,
  %``Towards the theory of the electroweak phase transition,''
  Phys.\ Rev.\  D {\bf 46}, 550 (1992).
%  [arXiv:hep-ph/9203203].
  %%CITATION = PHRVA,D46,550;%%
%}        


 \bibitem{sph-cond}   
%\MooreGE
%\lref\MooreGE{
  G.~D.~Moore,
  %``A nonperturbative measurement of the broken phase sphaleron rate,''
  Phys.\ Lett.\  B {\bf 439}, 357 (1998);
%  [arXiv:hep-ph/9801204].
  %%CITATION = PHLTA,B439,357;%%
%}
         
%
%\MooreSWA
%\lref\MooreSWA{
  G.~D.~Moore,
%``Measuring the broken phase sphaleron rate nonperturbatively,''
  Phys.\ Rev.\  D {\bf 59}, 014503 (1999).
%  [arXiv:hep-ph/9805264].
  %%CITATION = PHRVA,D59,014503;%%
%}

\bibitem{Ref:bsgNNLO}
 M.~Misiak and M.~Steinhauser,
 %``NNLO QCD corrections to the B -> X_s gamma matrix elements using
 %interpolation in m_c,''
 Nucl.\ Phys.\  B {\bf 764}, 62 (2007).
% [arXiv:hep-ph/0609241].

\bibitem{mg-2}
C. Amsler et al. (Particle Data Group), Physics Letters B667, 1 (2008).

        \bibitem{eg-2}
 P.~J.~Mohr, B.~N.~Taylor and D.~B.~Newell,
 %``CODATA Recommended Values of the Fundamental Physical Constants:
%2006,''
  Rev.\ Mod.\ Phys.\  {\bf 80}, 633 (2008).%;
%C. Amsler et al. (Particle Data Group), Physics Letters B667, 1 (2008).

 \bibitem{InvisibleDecayByDM}
  M.~C.~Bento, O.~Bertolami, R.~Rosenfeld and L.~Teodoro,
  %``Self-interacting dark matter and invisibly decaying Higgs,''
  Phys.\ Rev.\  D {\bf 62}, 041302 (2000);

 %  [arXiv:astro-ph/0003350].\bibitem{Bento:2001yk}
         M.~C.~Bento, O.~Bertolami and R.~Rosenfeld,
  %``Cosmological constraints on an invisibly decaying Higgs boson,''
  Phys.\ Lett.\  B {\bf 518}, 276 (2001).
%  [arXiv:hep-ph/0103340].
        
        
\bibitem{xmass}
 Y.~D.~Kim, Phys.\ Atom.\ Nucl.\ {\bf 69}, 1970 (2006).

\bibitem{Gustafsson:2007pc}
  M.~Gustafsson, E.~Lundstrom, L.~Bergstrom and J.~Edsjo,
  %``Significant gamma lines from inert Higgs dark matter,''
  Phys.\ Rev.\ Lett.\  {\bf 99}, 041301 (2007).
%  [arXiv:astro-ph/0703512].

\bibitem{Gehrels:1999ri}
  N.~Gehrels and P.~Michelson,
  %``GLAST: The next-generation high energy gamma-ray astronomy mission,''
  Astropart.\ Phys.\  {\bf 11}, 277 (1999).
  
\bibitem{Navarro:1996gj}
  J.~F.~Navarro, C.~S.~Frenk and S.~D.~M.~White,
  %``A Universal Density Profile from Hierarchical Clustering,''
  Astrophys.\ J.\  {\bf 490}, 493 (1997).
%  [arXiv:astro-ph/9611107].

\bibitem{nondec}
 
%\KanemuraVM
%\lref\KanemuraVM{
%  S.~Kanemura, et al.,
  S.~Kanemura, S.~Kiyoura, Y.~Okada, E.~Senaha and C.~P.~Yuan,
  %``New physics effect on the Higgs self-coupling,''
  Phys.\ Lett.\  B {\bf 558}, 157 (2003);

        %  [arXiv:hep-ph/0211308];
  %%CITATION = PHLTA,B558,157;%%
%}
%
%\KanemuraMG
%\lref\KanemuraMG{
  S.~Kanemura, Y.~Okada, E.~Senaha and C.~P.~Yuan,
  %``Higgs coupling constants as a probe of new physics,''
  Phys.\ Rev.\  D {\bf 70}, 115002 (2004).
%  [arXiv:hep-ph/0408364].
  %%CITATION = PHRVA,D70,115002;%%
%}

 \bibitem{hhh-measurement}
%\bibitem{eehhZ3}

C.~Castanier, P.~Gay, P.~Lutz and J.~Orloff, hep-ex/0101028; 

M.~Battaglia, E.~Boos and W.-M.~Yao, hep-ph/0111276;
%\YasuiSE
%\lref\YasuiSE{
%  Y.~Yasui, {\it et al.}, 
  Y.~Yasui, S.~Kanemura, S.~Kiyoura, K.~Odagiri, Y.~Okada, E.~Senaha and S.~Yamashita,
  %``Measurement of the Higgs self-coupling at JLC,''
  arXiv:hep-ph/0211047; 
  %%CITATION = HEP-PH/0211047;%%
%}
Talk given by S.~Yamashita at LCWS2004 
({http://polywww.in2p3.fr/actualites/congres/lcws2004/}). 


\bibitem{gamgamhh2}        

% \bibitem{jikia}
%\JikiaMT
%\lref\JikiaMT{
  G.~V.~Jikia,
%``Higgs boson pair production in high-energy photon-photon collisions,''
  Nucl.\ Phys.\  B {\bf 412}, 57 (1994);
%%CITATION = NUPHA,B412,57;%%
%}

%\bibitem{belusevic}
%\BelusevicPZ
%\lref\BelusevicPZ{
  R.~Belusevic and G.~Jikia,
  %``Higgs self-coupling in gamma gamma collisions,''
  Phys.\ Rev.\  D {\bf 70}, 073017 (2004).
%  [arXiv:hep-ph/0403303];
  %%CITATION = PHRVA,D70,073017;%%
%}
        
%\cite{Asakawa:2009ux}
%\bibitem{Asakawa:2009ux}
  E.~Asakawa, D.~Harada, S.~Kanemura, Y.~Okada and K.~Tsumura,
  %``Higgs boson pair production at the Photon Linear Collider in the two Higgs
  %doublet model,''
  arXiv:0902.2458 [hep-ph];
  %%CITATION = ARXIV:0902.2458;%%
        
%\cite{Takahashi:2009sf}
%\bibitem{Takahashi:2009sf}
  T.~Takahashi {\it et al.},
  %``Simulation Study of gamma gamma -> hh in a Photon Collider,''
  arXiv:0902.3377 [hep-ex].
  %%CITATION = ARXIV:0902.3377;%%

\bibitem{gamgamhh}

%\cite{Cornet:2008nq}
%\bibitem{Cornet:2008nq}
  F.~Cornet and W.~Hollik,
  %``Pair Production of Two-Higgs-Doublet_Model Light Higgs Bosons in $\gamma
  %\gamma$ Collisions,''
  Phys.\ Lett.\  B {\bf 669}, 58 (2008);
%  [arXiv:0808.0719 [hep-ph]];
  %%CITATION = PHLTA,B669,58;%%

%\cite{Asakawa:2008se}
%\bibitem{Asakawa:2008se}
  E.~Asakawa, D.~Harada, S.~Kanemura, Y.~Okada and K.~Tsumura,
  %``Higgs boson pair production at a photon-photon collision in the two Higgs
  %doublet model,''
  Phys.\ Lett.\  B {\bf 672}, 354 (2009);
%  [arXiv:0809.0094 [hep-ph]];
  %%CITATION = PHLTA,B672,354;%%

%\cite{Arhrib:2009gg}
%\bibitem{Arhrib:2009gg}
  A.~Arhrib, R.~Benbrik, C.~H.~Chen and R.~Santos,
  %``Neutral Higgs boson pair production in photon-photon annihilation in the
  %Two Higgs Doublet Model,''
  arXiv:0901.3380 [hep-ph];
  %%CITATION = ARXIV:0901.3380;%%

 \bibitem{wah}
%\KanemuraHZ
%\lref\KanemuraHZ{
  S.~Kanemura and C.~P.~Yuan,
  %``Testing supersymmetry in the associated production of CP-odd and  charged
  %Higgs bosons,''
  Phys.\ Lett.\  B {\bf 530}, 188 (2002);
%  [arXiv:hep-ph/0112165];
  %%CITATION = PHLTA,B530,188;%%
%}
         
                %\CaoTR
%\lref\CaoTR{
  Q.~H.~Cao, S.~Kanemura and C.~P.~Yuan,
  %``Associated production of CP-odd and charged Higgs bosons at hadron
  %colliders,''
  Phys.\ Rev.\  D {\bf 69}, 075008 (2004).
%  [arXiv:hep-ph/0311083].
  %%CITATION = PHRVA,D69,075008;%%
%}

\bibitem{hagiwara}
    
%\bullockFD
%\lref\BullockFD{
  B.~K.~Bullock, K.~Hagiwara and A.~D.~Martin,
  %``Tau polarization as a signal of charged Higgs bosons,''
  Phys.\ Rev.\ Lett.\  {\bf 67}, 3055 (1991).
  %%CITATION = PRLTA,67,3055;%%
%}

\bibitem{meg}
S.~Ritt,
 Nucl.\ Phys.\ Proc.\ Suppl.\ {\bf 162}, 279 (2006);
%

 %\bibitem{lfv-data}
%\cite{Baldini:2007zza}
%\bibitem{Baldini:2007zza}
   A.~Baldini,
   %``Status of the MEG experiment,''
   Nucl.\ Phys.\ Proc.\ Suppl.\  {\bf 168}, 334 (2007).
   %%CITATION = NUPHZ,168,334;%%

%\bibitem{haber-gunion-2hdm}
\bibitem{Ref:SMlike}
 J.~F.~Gunion and H.~E.~Haber,
 %``The CP-conserving two-Higgs-doublet model: The approach to the  decoupling
 %limit,''
 Phys.\ Rev.\  D {\bf 67}, 075019 (2003).
%  [arXiv:hep-ph/0207010].

\bibitem{Bertone:2004pz}
For a review, e.g.,
        G.~Bertone, D.~Hooper and J.~Silk,
  %``Particle dark matter: Evidence, candidates and constraints,''
  Phys.\ Rept.\  {\bf 405}, 279 (2005).
%  [arXiv:hep-ph/0404175].
        
 \bibitem{Bergstrom:1997fj}
  L.~Bergstrom, P.~Ullio and J.~H.~Buckley,
  %``Observability of gamma rays from dark matter neutralino annihilations  in
  %the Milky Way halo,''
  Astropart.\ Phys.\  {\bf 9}, 137 (1998).
%  [arXiv:astro-ph/9712318].

\bibitem{Cesarini:2003nr}
See e.g., A.~Cesarini, F.~Fucito, A.~Lionetto, A.~Morselli and P.~Ullio,
  %``The galactic center as a dark matter gamma-ray source,''
  Astropart.\ Phys.\  {\bf 21}, 267 (2004).
%  [arXiv:astro-ph/0305075].

\end{thebibliography}
\end{document}